\documentclass[a4paper,11pt]{article}
\usepackage[utf8]{inputenc}
\usepackage[table,xcdraw]{xcolor}
\usepackage{authblk}
\usepackage{graphicx}
\usepackage{newtxtext}
\usepackage{newtxmath}
\usepackage{hyperref}
\hypersetup{
    colorlinks = true,
    urlcolor   = blue,
    citecolor  = black,
}

\newcommand{\RomanNumeralCaps}[1]
\linenumbers

\title{Improved convergence of the spectral proper orthogonal decomposition through time shifting}

\author[1]{Diego C. P. Blanco \thanks{Correspondence address: diegodcpb@ita.br}}
\author[2]{Eduardo Martini}
\author[1]{Kenzo Sasaki}
\author[1]{André V. G. Cavalieri}

\affil[1]{Divisão de Engenharia Aeroespacial, Instituto Tecnológico de Aeronáutica, 12228-900, São José dos Campos/SP - Brazil}
\affil[2]{Département Fluides, Thermique et Combustion, Institut Pprime, CNRS, Université de Poitiers, ENSMA, 86000 Poitiers, France}

\date{}

\begin{document}
\maketitle

\begin{abstract}
Spectral proper orthogonal decomposition (SPOD) is an increasingly popular modal analysis method in the field of fluid dynamics due to its specific properties: a linear system forced with white noise should have SPOD modes identical to response modes from resolvent analysis. The SPOD, coupled with the  Welch method for spectral estimation, may require long time-resolved data sets. In this work, a linearised Ginzburg-Landau model is considered in order to study the method's convergence. SPOD modes of the white-noise forced equation are computed and compared with corresponding response resolvent modes. The quantified error is shown to be related to the time length of Welch blocks (spectral window size) normalised by a convective time. Subsequently, an algorithm based on a temporal data shift is devised to further improve SPOD convergence and is applied to the Ginzburg-Landau system. Next, its efficacy is demonstrated in a numerical database of a boundary layer subject to bypass transition. The proposed approach achieves substantial improvement in mode convergence with smaller spectral window sizes with respect to the standard method. Furthermore, SPOD modes display growing wall-normal and span-wise velocity components along the stream-wise direction, a feature which had not yet been observed and is also predicted by a global resolvent calculation. The shifting algorithm for the SPOD opens the possibility for using the method on datasets with time series of moderate duration, often produced by large simulations.
\end{abstract}

\section{Introduction}
\label{sec:intro}
Modal analysis is a well known method to study systems of partial differential equations (PDEs). By studying characteristics of a dynamic system, one can compute a basis in the PDE solution space and thus identify the underlying coherent structures governing the system's evolution in time and space. For instance, it is a established technique to develop reduced-order models of complex flows \cite{doi:10.1146/annurev-fluid-010816-060042,taira2020modal}.

In the context of dynamical systems, a popular method is the Proper Orthogonal Decomposition (POD), also called Principal Component Analysis (PCA) or Karhunen-Loève decomposition in other disciplines. Introduced to the fluid dynamics community by \cite{10012381873,lumley1970}, the statistical method decomposes an ensemble of bounded signals into orthogonal deterministic modes (often called empirical eigenfunctions) which, within all possible linear decompositions, are optimal in containing the most energy in an average sense \cite{berkooz_arfm_1993}.

The POD method enables the decomposition of velocity correlations from statistically stationary flows into modes, in order to identify and analyse the most energetic structures evolving in space and time. Due to its generality and solid foundation in the statistics theory, POD has been applied in numerous fluid dynamics fields, with studies involving coherent structures in turbulence \cite{sirovich1987}, derivations of reduced order models (ROMs) \cite{ravidran2000,noack_jfm_2003}, among many other applications. The method gained special traction in the 1990s with the rapid development of numerical analysis and computational tools \cite{Aubry1991}.

For statistical stationary flows, the application of POD over the Fourier transform in time of measured correlations generates modes that develop in a single frequency and display coherence in both space and time. This approach has been applied throughout the years to study jets \cite{glauser1987orthogonal,delville_ukeiley_cordier_bonnet_glauser_1999,picard_delville_2000,jung_gamard_george_2004,iqbal_thomas_2007,tinney_jordan_2008,gudmundsson_colonius_2011,schmidt_towne_colonius_cavalieri_jordan_bres_2017}, mixing layers \cite{arndt_long_glauser_1997,Bonnet1998,citriniti_george_2000}, turbulent pipes \cite{doi:10.1063/1.4902436}, boundary layers \cite{doi:10.1063/1.4974746}, wakes \cite{araya_colonius_dabiri_2017}.
	
This form of POD in the frequency domain has been labelled spectral POD, or SPOD \cite{towne_schmidt_colonius_2018,doi:10.2514/1.J058809}. It should be noted that we do not refer here to the homonym work of \cite{sieber_paschereit_oberleithner_2016}, which deals with a different algorithm. \cite{towne_schmidt_colonius_2018} established an important link between statistics and dynamical systems in the frequency space, as a linear set of equations, forced with an uncorrelated (spatially white) noise in a given frequency must have SPOD modes identical to the modes obtained via resolvent analysis. This property is not always present in the form of POD which yields purely spatial modes modulated by time dependent coefficients, hereby called spatial POD.

Later works employing the SPOD include: the development of data-driven reduced order models for boundary layer transition control \cite{sasaki_morra_cavalieri_hanifi_henningson_2020}, characterisation of noise generated over airfoils \cite{sano2019,abreu_tanarro_cavalieri_schlatter_vinuesa_hanifi_henningson_2021} and identification of coherent structures in channels \cite{ABREU2020108662}, wakes \cite{nidhan2020,doi:10.1063/5.0070092}, separated high-speed flows \cite{lugrin_beneddine_leclercq_garnier_bur_2021} and complex geometries \cite{doi:10.1063/5.0041395,doi:10.1063/5.0065929}.

Despite its versatility, the proposed SPOD algorithm \cite{towne_schmidt_colonius_2018} leads to some inherent complexity related to spectral estimation. It relies on the Welch method \cite{1161901} for the estimation of the Cross Spectral Density (CSD) tensor, and hence, computed gains and modes are sensitive to sampling parameters (number of field snapshots/realisations $N_t$, time step $\delta t$) and estimation parameters (number of blocks, $N_b$, number of snapshots per block, $N_{FFT}$, windowing function, $w(t)$ and overlap between successive blocks, $O_{FFT}$). For a given dataset with $N_t$ snapshots separated by a constant time step $\delta t$, increasing $N_b$ implies decreasing $N_{FFT}$ and \emph{vice-versa}. A larger $N_b$ penalises frequency resolution, increases spectral leakage and may lead to results contaminated with statistical bias, whereas a larger $N_{FFT}$ penalises convergence by decreasing the number of averaging blocks. The appropriate compromise between these cases is problem dependent and literature concerning which choice yields the best convergence for a given data set is lacking, even though statistical confidence bounds can be derived for computed SPOD gains \cite{doi:10.2514/1.J058809}. In general, estimation parameters are chosen subjectively based on the researcher's experience and convergence is checked \textit{a posteriori}.

Moreover, mode convergence can be slow and difficult to verify as the method requires long time-resolved data-sets. For Navier-Stokes simulations this translates to high computational and storage costs, that are often impracticable. A proposed alternative to reduce overall requirements of memory is the streaming SPOD algorithm proposed by \cite{SCHMIDT201998}. By accessing only one temporal snapshot at a time and incrementing the SPOD at each step, the algorithm allows for the use of arbitrarily long time series and real-time processing, avoiding storage costs, but maintaining long computations that may be expensive.

This work discusses the effects of estimation parameters on the convergence of the SPOD method. We first examine a model Ginzburg-Landau system which mimics the dynamics of turbulent jets. In sequence, the results from this convergence analysis are used to propose a new method based on a temporal shift to further reduce SPOD errors and enable a more accurate analysis of simulations with shorter integration times. In order to demonstrate its applicability to fluid flows, the shifting algorithm is applied to a boundary layer over a flat-plate subject to bypass transition.

For flat plate boundary layers, modal transition by Tollmien-Schlichting waves is bypassed for free-stream turbulence levels greater than around 1\% \cite{matsubara2001disturbance} as the transition occurs from the linear growth of stream-wise elongated, low-frequency, structures, with alternating high and low velocity, called streaks. The work of \cite{ellingsen1975stability} shows that perturbations in flows with low variation along the stream-wise direction can be algebraically amplified through the transient growth caused by the interaction of non-normal Orr-Sommerfeld modes, even in cases where modal stability predicts a exponential decay. This phenomenon is known in the literature as the lift-up effect \cite{brandt2014}. In bypass conditions, the streaks caused by the lift-up effect might be amplified enough to trigger non-linear interactions and degenerate to turbulence in Reynolds numbers below the critical levels provided by the linear theory of exponential growth \cite{brandt_schlatter_henningson_2004}.

Streaks appear as SPOD modes with vanishing frequency \cite{sasaki_morra_cavalieri_hanifi_henningson_2020}, and numerical convergence in such cases is challenging \cite{pickering_rigas_nogueira_cavalieri_schmidt_colonius_2020}. Hence, detection of coherent structures in a database of bypass transition of a boundary layer is a relevant test of convergence of SPOD modes and of the potential improvements that may be obtained with the proposed temporal shift of the database.

The remainder of the manuscript is organised as follows. Sections \ref{sec:SPOD} and \ref{sec:resolvent} review methods used in SPOD and resolvent analysis. In section \ref{sec:temporal_shift} the temporal shift is introduced and its effects are discussed, with guidance on how to choose shifting parameters. Finally, in section \ref{sec:application}, the results of the application of these concepts in the Ginzburg-Landau system and the flat plate case are evaluated.  

\section{Spectral Proper Orthogonal Decomposition}
\label{sec:SPOD}

Spectral Proper Orthogonal Decomposition is a modal analysis method aimed at extracting the most energetic turbulent coherent structures which evolve in a single frequency from statistically stationary data. A brief explanation of the algorithm is developed in this section. The computation involves the snapshot method, and further information can be found in \cite{towne_schmidt_colonius_2018} and \cite{doi:10.2514/1.J058809}.

Considering a turbulent flow data set with $M$ degrees of freedom (number of variables times number of spatial points $N_x$) and $N_t$ snapshots taken at a fixed interval $\delta t$ (each one treated as an independent realisation based on the ergodicity argument), we define the state vector $\mathbf{q} \in \mathbb{C}^{M}$ with a chosen inner product
\begin{equation}\label{eq:inner_product}
    \left\langle \mathbf{q}_{1}, \mathbf{q}_{2}\right\rangle_\mathbf{W}=\mathbf{q}_{1}^{H} \mathbf{W} \mathbf{q}_{2}
\end{equation}
where $\mathbf{W}^{M \times M}$ is a square positive definite diagonal matrix, and $\{\cdot\}^H$ is the conjugate transpose. The ensemble of realisation vectors are then used to construct the data matrix
\begin{equation} \label{eq:data_matrix}
    \mathbf{Q}=\left[\begin{array}{cccccc}
        \mid & \mid & & \mid \\
        \mathbf{q}_{1} & \mathbf{q}_{2} & \cdots & \mathbf{q}_{N_t} \\
        \mid & \mid & & \mid
    \end{array}\right], \quad \mathbf{Q} \in \mathbb{C}^{M \times N_t}
\end{equation}

For the case of an incompressible flow each component of the velocity field $\mathbf{u}(\mathbf{x},t)=\left(u(\mathbf{x},t),v(\mathbf{x},t),w(\mathbf{x},t)\right)$ is a variable. In that case, we can construct the vector $\mathbf{q} \in \mathbb{C}^{3 N_x}$ by computing the zero-mean velocity fluctuations for each snapshot,
\begin{equation} \label{eq:reynolds_decomp}
    \mathbf{\tilde{u}}(\mathbf{x},t) = \mathbf{u}(\mathbf{x},t) - \mathbf{U}(\mathbf{x}) = (\tilde{u}(\mathbf{x},t),\tilde{v}(\mathbf{x},t),\tilde{w}(\mathbf{x},t))
\end{equation} 
with $\mathbf{U}(\mathbf{x})$ being the mean field, and subsequently mapping all fluctuation components to a column vector in a convenient order, for example:
\begin{equation} \label{eq:state}
    \mathbf{q}_i = 
    \left[\begin{array}{ccccccccc}
        \tilde{u}(\mathbf{x_1},i) & \cdots & \tilde{u}(\mathbf{x}_{N_x},i) & \tilde{v}(\mathbf{x_1},i) & \cdots & \tilde{v}(\mathbf{x}_{N_x},i) & \tilde{w}(\mathbf{x_1},i) & \cdots & \tilde{w}(\mathbf{x}_{N_x},i)
    \end{array}\right]^T
\end{equation}

\subsection{Spectral estimation}

For the estimation of the CSD matrix $\mathbf{C}$, the Welch method is employed. The data matrix $\mathbf{Q}$ is decomposed into $N_b$ successive blocks containing $N_{FFT}$ realisations, with $O_{FFT}$ overlapped realisations between consecutive blocks. Then, the Discrete Fourier Transform (DFT) in time, weighted by a windowing function $w(t)$, is computed for each block.

Given $\mathbf{\hat{q}}_k(\omega)$ the vector resulting from the DFT at a frequency $\omega$ for the k-th block, we can assemble the vectors from all blocks corresponding to this frequency in a spectral data matrix  
\begin{equation}
    \mathbf{\hat{Q}}_\omega=\left[\begin{array}{cccccc}
        \mid & \mid & & \mid \\
        \mathbf{\hat{q}}_{1} & \mathbf{\hat{q}}_{2} & \cdots & \mathbf{\hat{q}}_{N_b} \\
        \mid & \mid & & \mid
    \end{array}\right], \quad \mathbf{\hat{Q}}_\omega \in \mathbb{C}^{M \times N_b}
\end{equation}
allowing the CSD matrix for each frequency to be computed as
\begin{equation}
    \mathbf{C}_\omega = \frac{1}{N_b} \mathbf{\hat{Q}}_\omega \mathbf{\hat{Q}}_\omega^{H} \mathbf{W}, \quad \mathbf{C}_\omega \in \mathbb{C}^{M \times M}
\end{equation}
and finally SPOD gains $\hat{\Lambda}$ and modes $\hat{\Psi}$ are calculated by solving the following eigenproblem of size $M \times M$:
\begin{equation} \label{eq:SPOD_eig}
    \mathbf{C}_\omega \hat{\Psi}=\hat{\Psi} \hat{\Lambda}
\end{equation}

\subsection{Snapshot SPOD}

In fluid dynamics applications, data sets often have $M \gg N_b$, as the problems are typicaly high dimensional and the cost of their simulation typically limits the amount of snapshots that can be obtained and handled. Therefore, it is computationally cheaper to obtain the same modes and gains by solving the much smaller eigenproblem of size $N_b \times N_b$ in the row space of the matrix $\mathbf{\hat{Q}}_\omega$, as presented by \cite{sirovich1987} and revisited in \cite{towne_schmidt_colonius_2018}:
\begin{equation}
    \mathbf{M}_\omega = \frac{1}{N_b} \mathbf{\hat{Q}}_\omega^{H} \mathbf{W} \mathbf{\hat{Q}}_\omega, \quad \mathbf{M}_\omega \in \mathbb{C}^{N_b \times N_b}
\end{equation}
\begin{equation}
    \mathbf{M}_\omega \hat{\Theta} = \hat{\Theta} \hat{\Lambda}
\end{equation}
\begin{equation}
    \hat{\Psi} = \frac{1}{\sqrt{N_b}} \mathbf{\hat{Q}}_\omega \hat{\Theta} \hat{\Lambda}^{-1/2}
\end{equation}

\section{Resolvent analysis}
\label{sec:resolvent}

When applying the Reynolds decomposition to the Navier-Stokes equation, linear and non-linear terms can be isolated. Then, non-linear terms may be treated as an input forcing signal to a linear system. In other words, we construct a state-space formulation with input/forcing terms $\mathbf{f}$ related to response/fluctuations terms $\mathbf{q}$ by a linear operator $\mathbf{L}$ which is only function of time invariant parameters, considering a linearisation of the Navier-Stokes system around a laminar solution \cite{jovanovic_bamieh_2005} or a mean flow \cite{mckeon_sharma_2010}. Operators $\mathbf{H}$ and $\mathbf{B}$ are used to restrict spatially the forcing and response, respectively, when necessary. We thus have
\begin{equation}
    \begin{aligned}
        \frac{\partial \mathbf{q}}{\partial t} &= \mathbf{L} \mathbf{q} + \mathbf{B}\mathbf{f} \\
        \mathbf{y} &= \mathbf{H} \mathbf{q}
    \end{aligned}
\end{equation}
and by taking the Fourier transform in time we obtain the Resolvent operator $\mathbf{R}$.
\begin{equation}
    \begin{aligned}
       \left(i \omega \mathbf{I} - \mathbf{L}\right) \mathbf{\hat{q}} &= \mathbf{B}\mathbf{\hat{f}}\\
        \mathbf{\hat{y}} &= \mathbf{H} \mathbf{\hat{q}}
    \end{aligned}
\end{equation}
\begin{equation} \label{eq: resolvent_op}
    \mathbf{\hat{y}} = \mathbf{R} \mathbf{\hat{f}} \implies \mathbf{R} = \mathbf{H}\left(i \omega \mathbf{I} - \mathbf{L}\right)^{-1}\mathbf{B}
\end{equation}

Now, resolvent response modes $\mathbf{U}$, forcing modes $\mathbf{V}$ and gains $\Sigma$ (ratio between the norms of $\mathbf{y}$ and $\mathbf{f}$), can be computed via the Singular Value Decomposition (SVD).
\begin{equation}\label{eq:svd_res}
    \mathbf{\tilde{R}} \mathbf{\tilde{V}} = \mathbf{\tilde{U}} \Sigma
\end{equation}

From the resolvent framework (eq. \ref{eq: resolvent_op}), the response CSD matrix $\mathbf{C}_\omega$ can be computed directly as a function of the forcing CSD matrix $\mathbf{F}_\omega$. Assuming

\begin{equation}
	\mathbf{C}_\omega = \mathcal{E}\left(\mathbf{\hat{y}} \mathbf{\hat{y}}^H\right), \quad \mathbf{F}_\omega = \mathcal{E}\left(\mathbf{\hat{f}} \mathbf{\hat{f}}^H\right)
\end{equation}

where $\mathcal{E}(\cdot)$ stands for the average over all realisations, we have
\begin{equation}
    \mathbf{C}_\omega = \mathbf{R} \mathbf{F}_\omega \mathbf{R}^H
\end{equation}

If forcing terms are perfectly uncorrelated, $\mathbf{F}_\omega = \mathbf{I}$, the expression can be reduced to 
\begin{equation}
    \mathbf{C}_\omega = \mathbf{R} \mathbf{R}^H 
\end{equation}
and, from eq. (\ref{eq:svd_res}), we have
\begin{equation} \label{eq:res_spod}
    \mathbf{C}_\omega = \mathbf{U} \Sigma^2 \mathbf{U}^{-1} \implies \mathbf{C}_\omega \mathbf{U} = \mathbf{U} \Sigma^2
\end{equation}
leading to
\begin{equation}
    \hat{\Psi} = \mathbf{U}, \quad \hat{\Lambda} = \Sigma^2
\end{equation}
from eq. (\ref{eq:SPOD_eig}).

This shows that resolvent response modes are equal to those obtained via the SPOD method for a system forced with spatially-white forcing \cite{towne_schmidt_colonius_2018}, a property also true for an arbitrary inner product norm (see Appendix \ref{appA}).

In this study, the mode equivalence property is exploited to study numerically the convergence of SPOD modes by employing a simple linear system analogous to the Navier-Stokes system, forced with spatially uncorrelated signals, and comparing resolvent and SPOD results.

\section{Temporal shift}
\label{sec:temporal_shift}

The idea of introducing a temporal shift to compensate for transport in signal processing and modal decomposition is not new. The work of \cite{doi:10.1137/17M1140571}, for instance, discusses its application on the construction of ROMs and proposes a shifted method based on spatial POD to address the slow decay of energy observed from the modal decomposition of transport dominated phenomena. However, to the best of our knowledge the application of a time delay in SPOD has not been attempted.

\begin{figure}
    \centering
    \includegraphics[width=\linewidth]{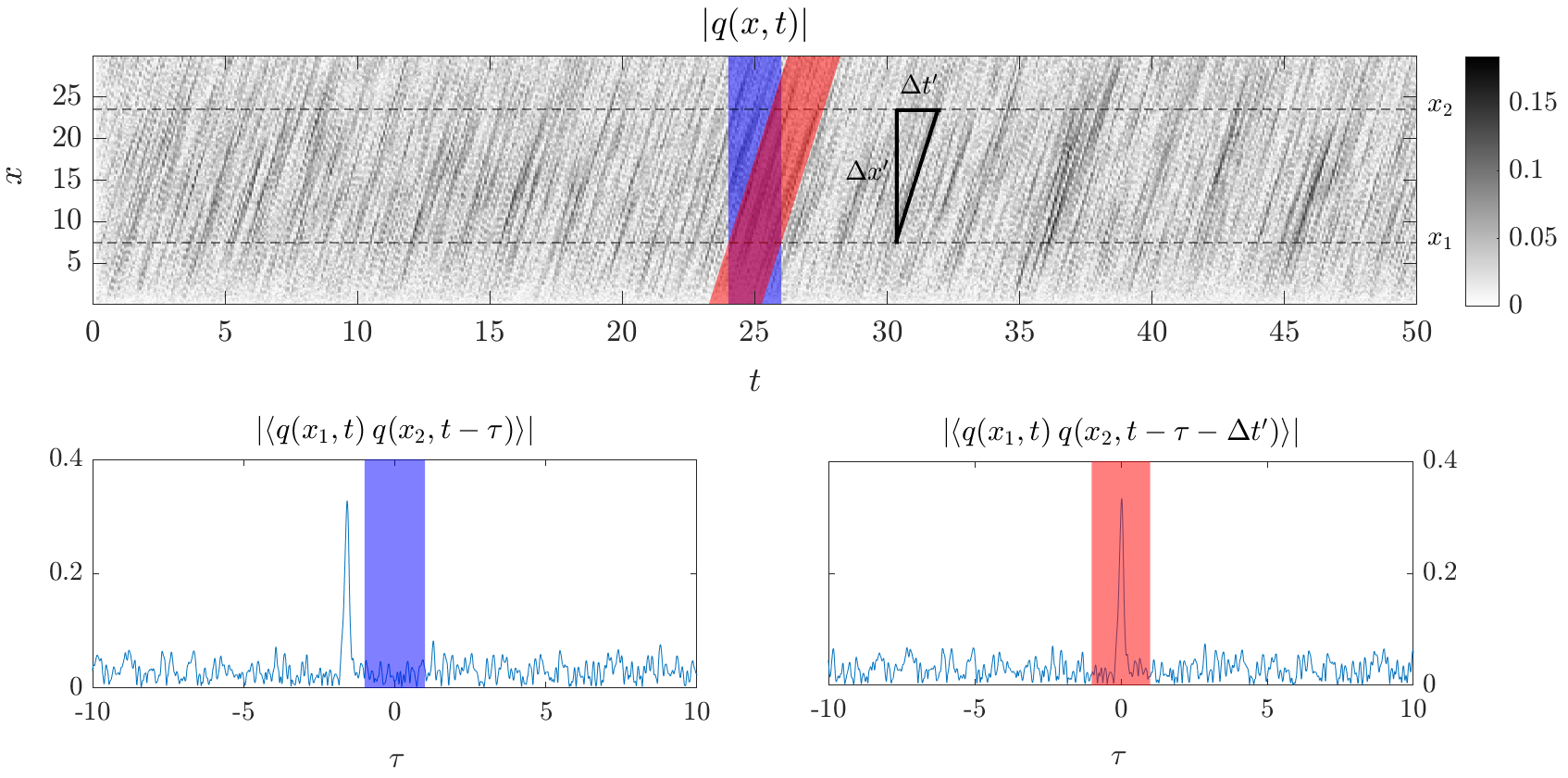}
    \caption{Diagram illustrating the effect of the temporal shift on the model system presented in section \ref{sec:GL}. In standard conditions, the cross-correlation peak lies outside the chosen time window. The application of the expected time lag between signals in positions $x_1$ and $x_2$ maintains the peak inside the same time window. The operator $\langle \cdot \rangle$ denotes expected value.}
    \label{fig:diagram}
\end{figure}

The Welch method employed in the SPOD method separates the data set in blocks of a chosen time length. For transport dominated phenomena with elongated domains, one can usually choose signals corresponding to coordinates $\mathbf{x_1}$ and $\mathbf{x_2}$ where the time lag for peak cross-correlation is greater than the block time length. The effect is shown in figure \ref{fig:diagram} with the blue rectangle, which illustrates the standard situation where the same initial and final times of Welch blocks are taken for all positions in the flow. This causes an apparent loss of coherence and encumbers the statistical convergence of the decomposition. This feature was observed, for instance, by \cite{PhysRevFluids.2.024604}.

\subsection{Proposed algorithm}

In order to address the limitations imposed by the Welch method, we propose an extension to the SPOD algorithm in two steps, exploiting the assumed stationarity: we first apply a shift in time and then correct the phase in Fourier space, in a procedure that preserves asymptotic convergence to the true CSD (see Appendix \ref{appB}). The temporal shift aims at improving convergence for a window of finite size by maximising cross-correlations within each Welch block, via the insertion of an expected time delay to each spatial location. 

Concerning the SPOD implementation, first we treat the data matrix $\mathbf{Q}$ by means of a data shift and then correct the effects of this treatment after the application of the Welch method. Given a point of interest with stream-wise coordinate $x_p$ and a matrix $\mathbf{Q}$ with $N_x$ rows and $N_t$ columns,  for each row $k$ the columns are shifted by 
\begin{equation} \label{eq:shift}
    s_k = \frac{\Delta x}{U_k \delta t}, \quad \Delta x = x_p-x_k
\end{equation}
positions. In this expression $x_k$ is the point's stream-wise coordinate, $\delta t$ is the time step between realisations and $U_k$ is a chosen velocity so that $\Delta x /U_k$ approximates the time lag for peak cross-correlation with respect to the point of interest. For instance, the dominant convection velocity could be used, if this information is readily available. Moreover, there is no requirement for $U_k$ to be constant throughout the domain.  

Columns at the beginning or the end of the time series, for which the time shift $s_k$ is not possible, are removed from the computation. This can be more easily implemented as a circular shift of the data removing the columns for which there is an artificial overlap between the end of the time series at one point and the beginning of the series at another point.

It is important to notice that, if $s_k$ is not an integer, the shift must be interpolated, as discussed in \cite{doi:10.1137/17M1140571}. Given $\mathcal{Z}^s_k$ the operator that circularly shifts a line $k$ by $s$ positions and $\mathbf{\tilde{Q}}$ the resulting shifted matrix, the linearly interpolated shifting operation takes the form
\begin{equation}
    \mathbf{\tilde{Q}} = \left\{
    \begin{array}{ll}
        \mathcal{Z}^{s_k}_k \mathbf{Q}, & \text{if} \: s_k \in \mathbb{Z} \\
        \left(\lceil s_k \rceil - s_k\right) \cdot \mathcal{Z}^{\lfloor s_k \rfloor}_k \mathbf{Q} + \left(s_k - \lfloor s_k \rfloor\right) \cdot \mathcal{Z}^{\lceil s_k \rceil}_k \mathbf{Q}, & \text{otherwise}
    \end{array}\right.
\end{equation}

After application of the Welch method the k-th row of each resulting $\mathbf{\hat{Q}}_\omega$ matrix, with $N_x$ rows and $N_b$ columns, is multiplied by $\exp(-i \omega s_k \delta t)$ to correct the phase created by the previous temporal shift operation, according to the properties of the Fourier transform. 

This procedure will be referred in this work as shifted SPOD, for the sake of concision, and should not be considered a new variety of POD, but instead a new algorithm for the method we call SPOD.

Computationally, the shifting algorithm deals with row-wise rearrangements of matrix elements and also row-wise array multiplications, which are low complexity operations when compared to Fourier transforms and eigenproblems. Once the data matrices are loaded in memory, both these steps should be rapidly performed in a modern desktop computer without the addition of expressive cost to the SPOD method.

\subsection{On the choice of point of interest, shifting velocity and window size} \label{sec:choice}

The choice of window size $N_{FFT}$ and, consequently, the number of blocks $N_b$ is the most subjective step of the SPOD method. For a given database, it is not obvious how one should choose the most appropriate number of realisations per block that simultaneously ensures blocks are long enough in time to properly capture the evolution of the most energetic structures and numerous enough to reduce the decomposition's variance.

Here, we discuss this choice and propose a pre-processing analysis to objectively set a suitable $N_{FFT}$ for a given dataset, considering a shifting velocity and point of interest. This procedure is also applicable to the standard SPOD, as it can be treated as a special case of the shifted SPOD with shifting velocity $U_k \to \infty$. The procedure follows four steps:
\begin{enumerate}
    \item Given a preferential flow direction $X$, we choose a point of interest $p$ of coordinate $x_p$ contained in the region from which we want to extract the most energetic structures. For simulations, it is important to avoid points under the influence of numerical boundaries, such as inlets, outlets and sponge/fringe zones, in order to guarantee that computed two-point correlations are physical.
    \item Once $p$ is chosen, we compute the space-time correlations at positions along the $X$ direction with respect to the signal at $x_p$, as later illustrated in figure \ref{fig:GL_xcorr}(left). In the case of multi-dimensional data, cross-correlations can be computed at multiple slices, as will be later shown in figure \ref{fig:BL_xcorr}(a,c,e), or for multiple vector components.  
    \item The inclination of the peak correlations quantifies the dominant convection velocity, which might vary as a function of position and not necessarily match the local mean velocity \cite{doi:10.1146/annurev-fluid-010816-060309}. By applying a shifting velocity that approximates said dominant velocity, it is possible to align correlation peaks, as later discussed in figures \ref{fig:GL_xcorr}(right) and \ref{fig:BL_xcorr}(b,d,f), in order to generate a vertical band of high correlation in time. This narrower correlation allows a given window to capture more kinematic information about the system. However, if a dominant velocity cannot be identified, the shifting operation should yield no special benefit.
    \item From the space-time plot we are finally able to determine the smallest window size capable of capturing the bulk of correlation, which simultaneously resolves the frequencies of interest in the study. By narrowing the correlation band in time, the shifting algorithm admits even smaller window sizes for the same amount of correlation captured and thus reduces overall variance by increasing the number of blocks $N_b$. In this way, it becomes feasible to objectively set the $N_{FFT}$ parameter based on the individual characteristics of each database and frequency resolution requirements.
\end{enumerate}

\section{Convergence analysis of SPOD}
\label{sec:application}

\subsection{A model problem: Ginzburg-Landau equation}
\label{sec:GL}

The Complex-valued Ginzburg-Landau Equation (CGLE) has been one of the most studied equations in a wide variety of physics fields to provide insight over the dynamics of non-equilibrium phenomena in spatially extended systems \cite{RevModPhys.74.99}. Particularly in the field of fluid dynamics and turbulence, the one-dimensional linearised complex Ginzburg-Landau equation is often employed to model instabilities in spatially evolving flows for its properties of advection and diffusion \cite{doi:10.1146/annurev.fl.22.010190.002353,bagheri2009input,10.1115/1.4042736}. The equation is written in the input-output framework as
\begin{equation}
    \frac{\partial \mathbf{q}}{\partial t} + U \frac{\partial \mathbf{q}}{\partial x} - \mu(x)\mathbf{q} - \gamma \frac{\partial^2 \mathbf{q}}{\partial x^2} = \mathbf{f},	
\end{equation} 
with $\mathbf{q}$ and $\mathbf{f}$ being output response and input forcing respectively, and rewritten in Fourier space as
\begin{equation}
    \left(i \omega \mathbf{I} + U \frac{\partial}{\partial x} - \mu(x) - \gamma \frac{\partial^2}{\partial x^2}\right) \mathbf{\hat{q}} = \mathbf{\hat{f}}	
\end{equation} 
from which the Resolvent operator can be computed as
\begin{equation}
    \mathbf{R}	=  \left(i \omega \mathbf{I} + U \frac{\partial}{\partial x} - \mu(x) - \gamma \frac{\partial^2}{\partial x^2}\right)^{-1}
\end{equation}
with an inverse obtained after a discretisation of the derivatives to form the operators in matrix form.

When the system is forced with a spatially white statistics on a given frequency $\omega$, the SPOD modes $\mathbf{\hat{\psi}}(\omega)$ are identical to the  resolvent response modes, $\mathbf{u}(\omega)$. Thus, such response modes, which are deterministic in nature, are used to quantify the SPOD convergence defined in terms of the error
\begin{equation}\label{eq:epsilon}
    \varepsilon(\omega) = 1 - \frac{| \langle \mathbf{\hat{\psi}}(\omega),\mathbf{u}(\omega) \rangle|}{||\mathbf{\hat{\psi}}(\omega)|| \cdot ||\mathbf{u}(\omega)||} 
\end{equation}
where the operators $\langle .,. \rangle$ and $||.||$ are respectively the euclidean inner product and norm.   

The model problem was devised according to \cite{10.1115/1.4042736} and parameters
\begin{equation}
    \mu (x) = A\left(1-\frac{x}{10}\right)
\end{equation}
\begin{equation}
    \gamma = \frac{1-i}{10}
\end{equation} 
were set to mimic the dynamics of turbulent jets. Three distinct cases were investigated: case 1 with $U=10$ and $A=0.6$; case 2 with $U=12$ and $A=1$; case 3 with $U=14$ and $A=1.25$. These choices lead to a varying convection velocity between the three cases, which modifies the correlation peaks discussed in section \ref{sec:temporal_shift}. Numerical solutions were obtained using a Crank-Nicolson scheme for time marching. Spatial discretisation was set using a fixed grid spacing of $\Delta x = 0.1$, with a total of $N_x = 300$ points, $x \in [0,30]$ and $q(0)=q(30)=0$. A second-order upwind scheme was applied for the first spatial derivative and a second-order centred scheme for the second. Because of the uniform grid, the inner-product matrix $\mathbf{W}$ is set as $I\Delta x$ to account for the spatial quadrature. 

The forcing is applied by adding a random complex number with uniformly distributed phase and amplitude to the value at each grid point for every time step. To ensure that fluctuations are adequately resolved in frequency, we apply to the forcing signal a finite impulse response (FIR) low-pass filter of 30th order in time, with cut-off at 60\% of the Nyquist frequency. Numerical integration is done for a range of 500 time-units using a time step of $\delta t = 0.01$, generating a total of $N_t = 5\times 10^4$ snapshots per case.

\subsubsection{Computation of SPOD modes}
\label{sec:converg_GL}

In order to understand how SPOD errors scale as a function of sampling and estimation parameters for each model case, the number of snapshots $N_{FFT}$ within each block is varied from 100 to 5000 in steps of 100 and the number of blocks $N_b$ set accordingly to comprise all $N_t$ snapshots (fixed-data analysis). This configuration ensures that all computations reach exactly the frequency $\omega=2\pi$, which has the same order of magnitude of the most energetic frequency in all three cases, in order to compare SPOD and Resolvent modes. 

To reduce spectral leakage effects and allow for the use of a relatively large consecutive block overlapping of $O_{FFT} = \lfloor 0.75 \: N_{FFT} \rfloor$, we apply to each block an infinitely-smooth windowing function of the form
\begin{equation} \label{eq:wind_func}
    w_{C^{\infty}_2}(t)=\left\{\begin{array}{ll}
        \mathrm{e}^{8} / \mathrm{e}^{\frac{2 T^{2}}{t(T-t)}} & , \: 0<t<T \\
        0 & , \text { otherwise }
    \end{array}\right.
\end{equation} 
with all derivatives equal to zero at $0$ and $T$ \cite{martini2020accurate}.

\begin{figure}
    \centering
    \includegraphics[width=0.9\linewidth]{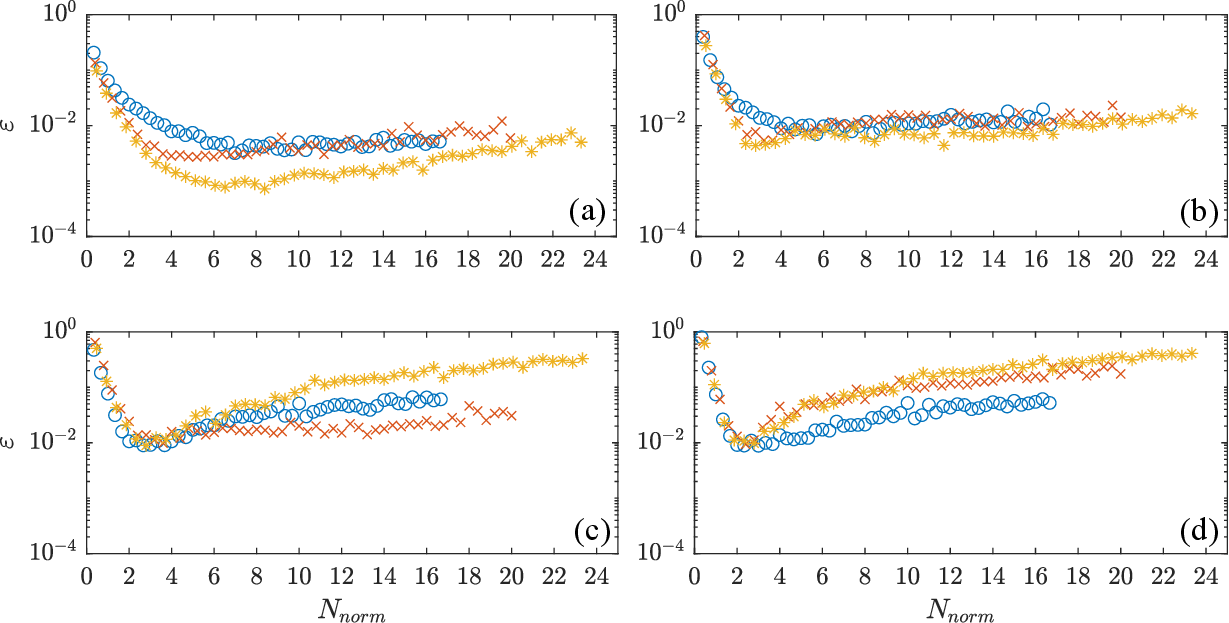}
    \caption{SPOD estimation error $\varepsilon$ at frequency $\omega=2 \pi$, plotted against the normalised window length $N_{norm}$. Legend: ($\color{blue}\circ$) Case 1; ($\color{red}\times$) Case 2; ($\color{orange}*$) Case 3. (a) Mode 1; (b) Mode 2; (c) Mode 3; (d) Mode 4.}
    \label{fig:errors}
\end{figure}

\begin{figure}
    \centering
    \includegraphics[width=0.88\linewidth]{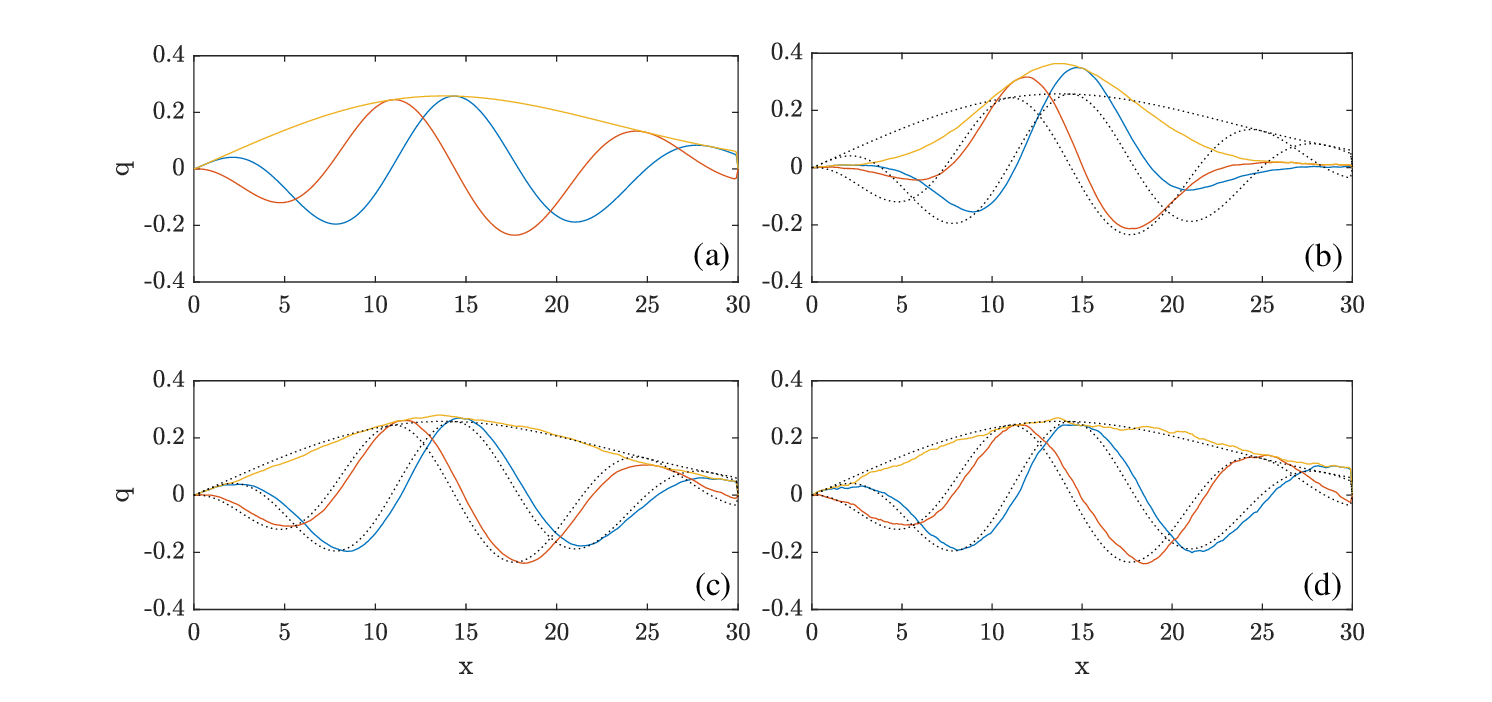}
    \caption{Mode 1 at frequency $\omega=2 \pi$, for case 3. Other cases display similar features. (a) Resolvent operator response mode; (b) SPOD mode with $N_{FFT}=100$, $N_b=1997$ and $N_{norm} \approx 0.5$: higher precision, lower accuracy (bias); (b) SPOD mode with $N_{FFT}=600$, $N_b=330$ and $N_{norm} \approx 3$; (c) SPOD mode with $N_{FFT}=5000$, $N_b=37$ and $N_{norm} \approx 23$: lower precision, higher accuracy (noise).\\ Legend: ($\color{orange}\blacksquare$) Absolute value; ($\color{blue}\blacksquare$) Real part; ($\color{red}\blacksquare$) Imaginary part; (Dotted lines) Resolvent mode for comparison.}
    \label{fig:bias_noise}
\end{figure}

The error between SPOD and resolvent, defined in eq. (\ref{eq:epsilon}), is shown in figure \ref{fig:errors} for the leading four SPOD modes at frequency $\omega=2 \pi$. The results obtained indicate that better absolute convergence is related to window length, normalised by domain size, and convection velocity. Thus, we define a normalised window size
\begin{equation} \label{eq:n_norm}
    N_{norm} = \frac{N_{FFT} \: \delta t}{T_c} , \quad T_c = \frac{L}{U} 
\end{equation}
where $T_c$ is the time taken to cross the entire domain based on the parameter $U$ of the CGLE system and $L$ is the total length of the domain. 

The graphs in figure \ref{fig:errors} show that SPOD modes computed with very low values of $N_{norm}$ yield the worst convergence. In that scenario, the Welch blocks are too short to correctly capture the evolution of coherent structures in time, which leads to statistically biased results. On the other side, for larger $N_{norm}$ values, errors tend to stabilise or even increase for higher modes (modes 3 and 4) due to the reduced number of averaging blocks and consequent increase in noise.

A trade-off between estimation bias, which is reduced with increasing window sizes, and estimation variance, i.e., statistical noise, which increases with window size (lower number of samples) is thus required. Effects of bias and noise are illustrated in figure \ref{fig:bias_noise}. 

Overall, a normalised window size between 2 and 4 is a good compromise to achieve the best convergence across the first 4 SPOD modes in these cases. This implies that a time series should be sufficiently long to comprise a large number of blocks of normalised window size between 2 and 4. This would translate to a database with several flow-through times, which may be quite expensive to compute and store if one deals with complex flows. In what follows we will explore how the temporal shift of section \ref{sec:temporal_shift} may alleviate such requirements.  

\subsubsection{Application of the temporal shift}
\label{sec:shift_GL}

\begin{figure}
    \centering
    \includegraphics[width=0.9\linewidth]{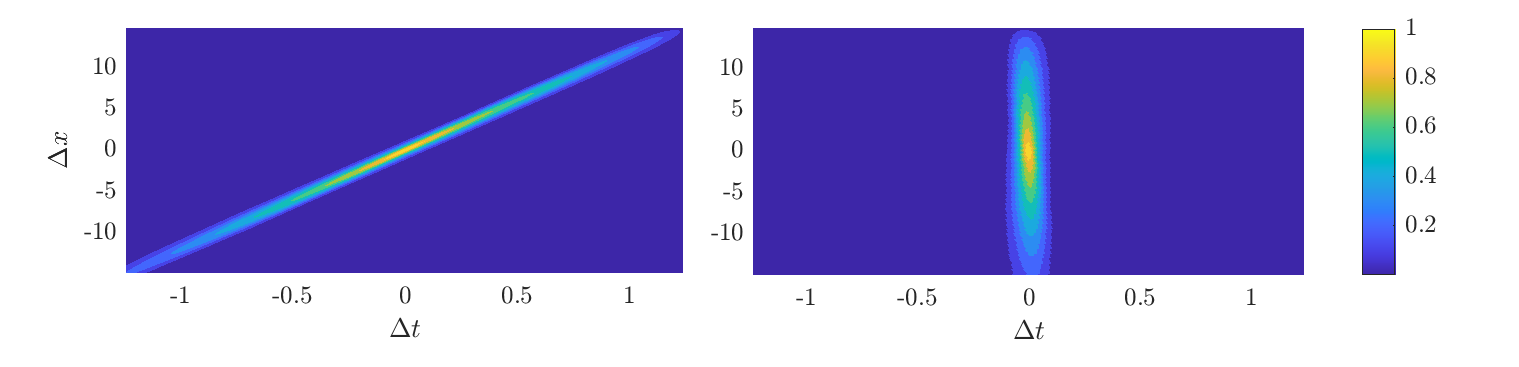}
    \caption{Space-time cross-correlation with respect to the point $x_p=15$ for case 2 ($U=12$, $A=1$). Other cases display similar features. (Left) No shift; (Right) Optimal shift.}
    \label{fig:GL_xcorr}
\end{figure}

The analysis of section \ref{sec:converg_GL} was also performed using the shifted SPOD algorithm. Parameters $N_{FFT}$, $N_b$ and $O_{FFT}$ were set in the same exact manner, as well as the windowing function. Each case was shifted with respect to the point of interest $x_p=15$, in the middle of the numerical domain, according to the algorithm of section \ref{sec:temporal_shift}. The parameter $U$ of the corresponding CGLE was applied as the optimal shifting velocity for all points, in a operation that approximates almost exactly the peak cross-correlation, as it is shown in figure \ref{fig:GL_xcorr}. In this graph a Welch block window can be represented as a vertical band, consisting of all $\Delta x$ and a limited $\Delta t$. From this observation is possible to conclude that a narrow window can contain more correlation in the shifted case.

\begin{figure}
    \centering
    \includegraphics[width=0.9\linewidth]{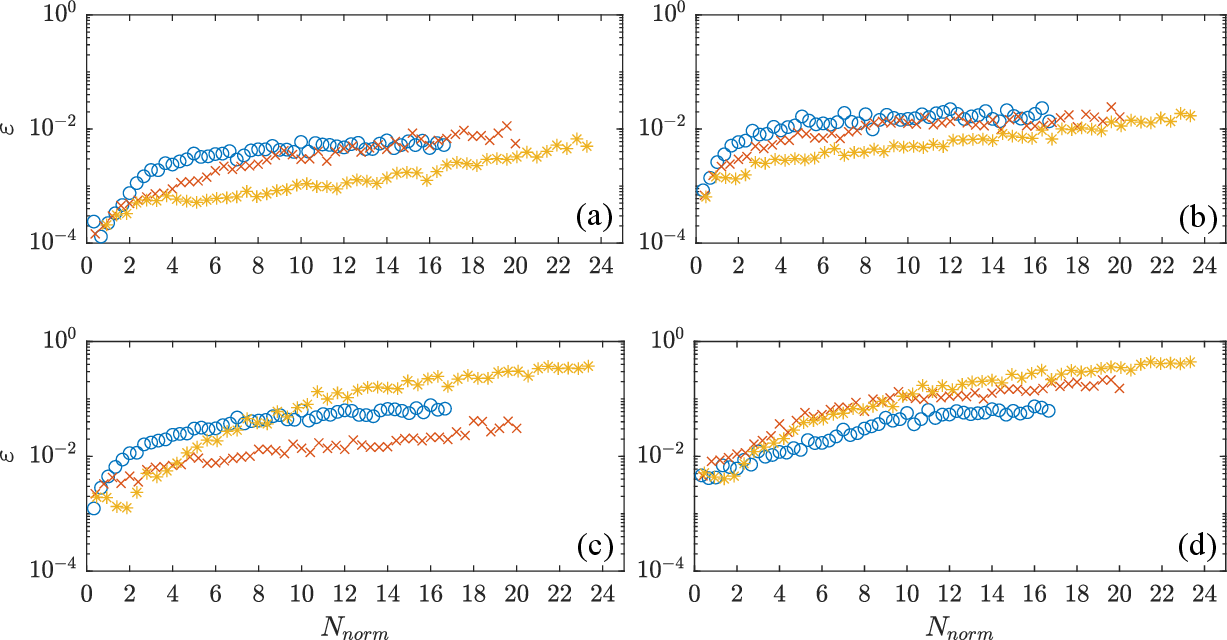}
    \caption{Shifted SPOD estimation error $\varepsilon$ at frequency $\omega=2 \pi$, plotted as a function as $N_{norm}$. Legend: ($\color{blue}\circ$) Case 1; ($\color{red}\times$) Case 2; ($\color{orange}*$) Case 3. (a) Mode 1; (b) Mode 2; (c) Mode 3; (d) Mode 4.}
    \label{fig:errors_shifted}
\end{figure}

\begin{figure}
    \centering
    \includegraphics[width=\linewidth]{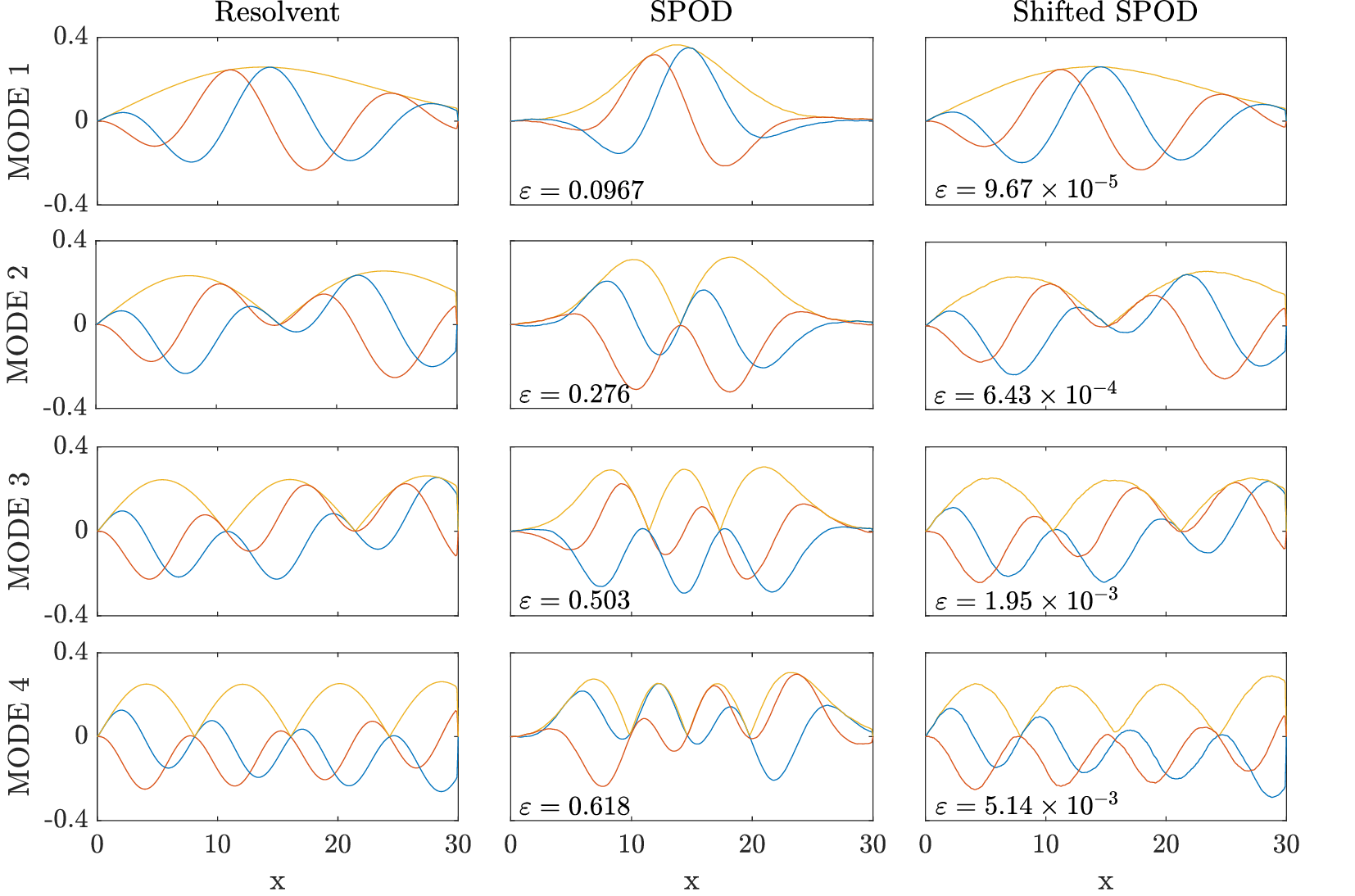}
    \caption{Comparison between Resolvent, SPOD and Shifted SPOD modes, at frequency $\omega=2 \pi$, for case 3. All SPOD modes were computed with $N_{FFT}=100$ and $N_{norm} \approx 0.5$. Other cases display similar features.\\ Legend: ($\color{orange}\blacksquare$) Absolute value; ($\color{blue}\blacksquare$) Real part; ($\color{red}\blacksquare$) Imaginary part.}
    \label{fig:compare}
\end{figure}

The results obtained are shown in figure \ref{fig:errors_shifted}. Compared to figure \ref{fig:errors}, estimation errors $\varepsilon$ are greatly reduced for the lowest values of $N_{norm}$, which means that the shifting operation is successful in reducing the bias observed with shorter Welch blocks, an effect illustrated in figure \ref{fig:compare}. Thus, the approach considerably alleviates the trade-off between bias and variance discussed in the previous section.

One should note that, in these cases, where the correlation between signals falls sharply for time lags which do not match the convection velocity (figure \ref{fig:GL_xcorr}), the minimal time window able to resolve the studied frequency is successful in reducing the bias. However, it is expected that, in more general cases, containing a wide range of coherent structures and non-linear dynamics, the narrowest time windows able to resolve a given frequency should still lead to errors, as they might not contain a relevant part of the correlation function.

\subsubsection{Comparison with fixed variance}

We compare standard and shifted SPOD algorithms when a fixed number of blocks $N_b$ is considered and only the window size $N_{FFT}$ is changed. Contrary to the previous analysis, this setup does not consider all of the snapshots available in the data set, as the SPOD algorithm will require fewer snapshots for smaller window sizes and the same quantity of blocks. The variance of the estimates is kept approximative constant, allowing for a study of the effect of estimation bias. We choose $N_b = 100$ and sweep $N_{FFT}$ values from 100 to 1900 in steps of 100. The overlap value $O_{FFT}$ and windowing function $w(t)$ remain unchanged. 

\begin{figure}
	\centering
	\includegraphics[width=\linewidth]{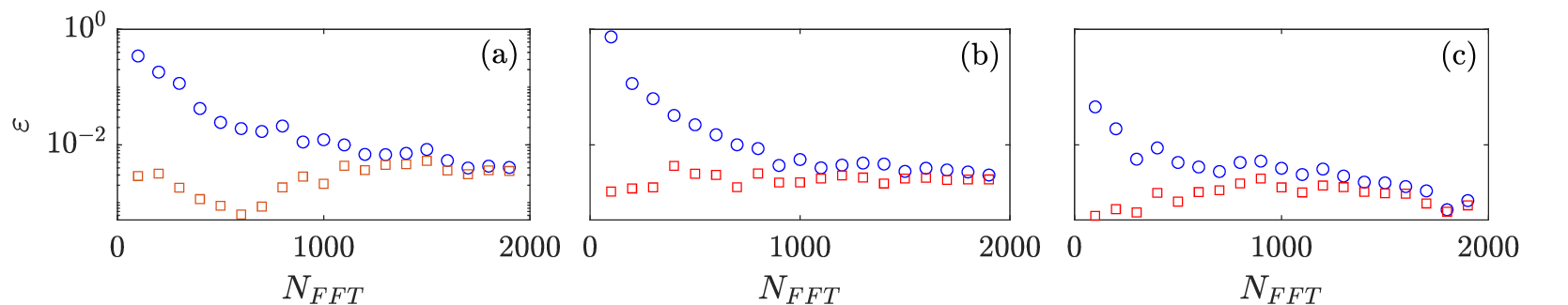}
	\caption{Estimation error $\varepsilon$ of mode 1, at frequency $\omega=2 \pi$ and number of blocks $N_b=100$, as function of window size ${N_{FFT}}$.\\ Legend: ($\color{blue}\circ$) Standard SPOD; ($\color{red}\square$) Shifted SPOD. (a) Case 1; (b) Case 2; (c) Case 3.}
	\label{fig:mode1_fixedBlocks}
\end{figure}

Figure \ref{fig:mode1_fixedBlocks} shows the results of this fixed-blocks analysis. Estimation errors for the Shifted SPOD are always smaller than those of standard SPOD, with effects more pronounced in the region of smaller window size $N_{FFT}$, where a reduction of at least one order of magnitude is observed in some cases.

An expressive reduction in estimation error for smaller $N_{FFT}$ values induced by the shifted SPOD algorithm is relevant because, for a given number of Welch blocks, it allows a more precise and accurate computation of most energetic modes on datasets with short time lengths, as is typical for large simulations, and reduces the sensitivity of the method to the window size parameter.

\subsubsection{Comparison with fixed window size}

We also compare standard and shifted SPOD algorithms considering a fixed windows size $N_{FFT}$ where only the number of blocks $N_b$ is changed. As in the previous analysis, this setup does not consider all of the snapshots available in the data set. We choose $N_{FFT} = 100$, which corresponds to $N_{norm} < 1$ for all three cases, and sweep $N_{b}$ values from 20 to 800 in steps of 20. Again, the overlap value $O_{FFT}$ and windowing function $w(t)$ remain unchanged.

\begin{figure}
    \centering
    \includegraphics[width=\linewidth]{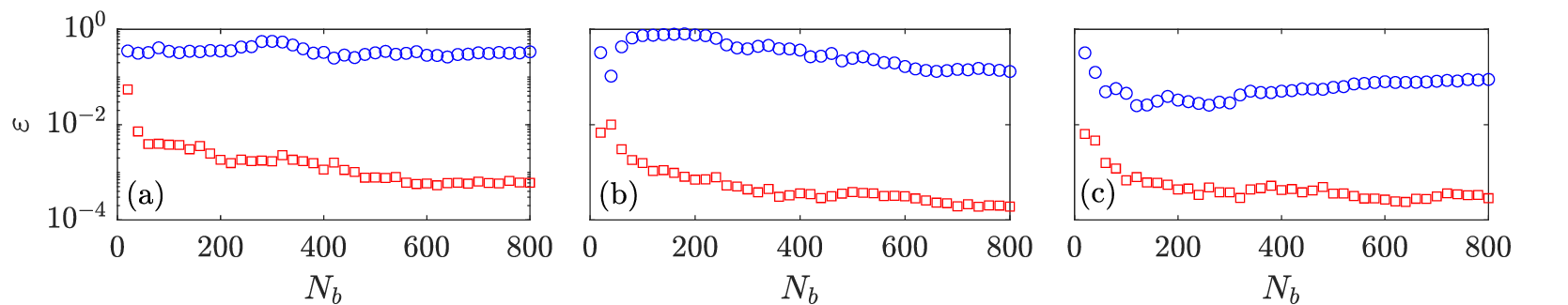}
    \caption{Estimation error $\varepsilon$ of mode 1, at frequency $\omega=2 \pi$ and number of snapshots $N_{FFT}=100$, as function of number of blocks ${N_{b}}$.\\ Legend: ($\color{blue}\circ$) Standard SPOD; ($\color{red}\square$) Shifted SPOD. (a) Case 1 (b) Case 2 (c) Case 3.}
    \label{fig:mode1_fixedWindow}
\end{figure}

Figure \ref{fig:mode1_fixedWindow} shows the results of the fixed-window analysis. For the case of a small windows size, not long enough in time to properly capture the system dynamics, the standard SPOD algorithm is not capable of reducing estimations errors with the increase of the time series length (and consequently the number of blocks) to the level of the shifted algorithm, which shows significantly lower errors. 

\subsubsection{Effect of non-optimal shifting velocities}

All previous analysis focused on the optimal shift, equal to the known single convection velocity in each case. However, in most fluid dynamics applications, different scales propagate at different velocities. Moreover, as stated in section \ref{sec:choice}, the convection of coherent structures might vary as a function of position and not necessarily match the local mean velocity, making the exact correspondence between shifting and convection velocities impractical. 

\begin{figure}
	\centering
	\includegraphics[width=\linewidth]{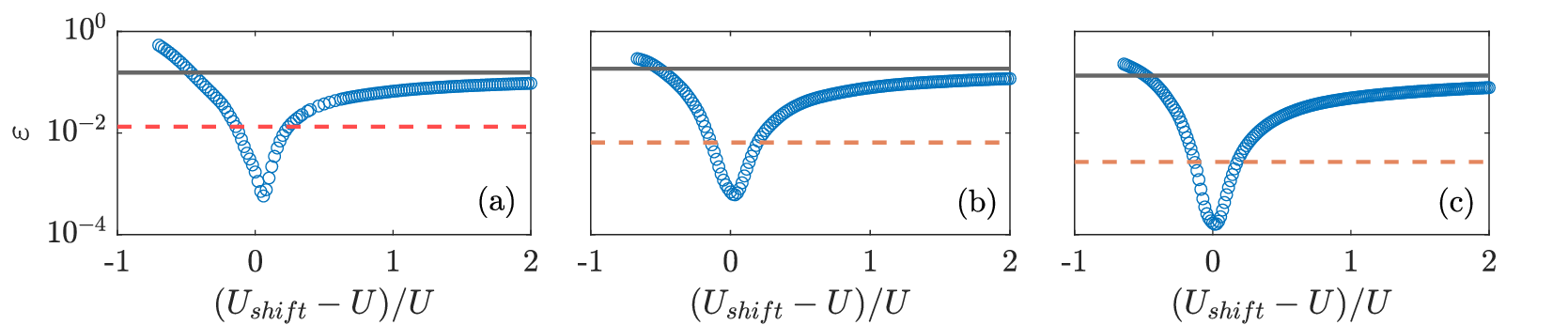}
	\caption{Estimation error $\varepsilon$ of mode 1, at frequency $\omega=2 \pi$, number of snapshots $N_{FFT}=100$ and entire time series, as function of shifting velocity $U_{shift}$.\\ Legend: ($-$) Standard SPOD; ($\color{red} - - $) Random $U_{shift}(x) \in [0.8 \: U, 1.2 \: U]$ shifted SPOD; ($\color{blue}\circ$) Constant $U_{shift}$ shifted SPOD. (a) Case 1 (b) Case 2 (c) Case 3.}
	\label{fig:mode1_changingU}
\end{figure}

To study the effects of a non-optimal shift, we compare the errors $\varepsilon$ of shifted SPOD modes computed with different shifting velocities $U_{shift}$. Figure \ref{fig:mode1_changingU} shows the results of this analysis, where it is clear that the lowest error is achieved when shifting and convection velocities are equal.
    
According to eq. (\ref{eq:shift}), the shift is inversely proportional to shifting velocity. For smaller than optimal $U_{shift}$, the algorithm overestimates the necessary shift and errors rise sharply, as $U_{shift} \to 0$ and $s_k \to \infty$, reaching levels higher than the standard SPOD. Due to this behaviour, very low shifting velocities should be applied with special care, even though the use of very low convection velocities is rare, as quantities are often normalised. 
    
Inversely, $U_{shift} \to \infty$ leads to $s_k \to 0$, implying shifted and standard SPODs are asymptotically identical, with errors following this same trend. In other words, assuming errors grow monotonically with the increase of $U_{shift}-U$, the shifted SPOD will always return better converged results than the standard algorithm for larger than optimal $U_{shift}$. This property can be exploited in the case of a flow with multiple known convection scales propagating to the same direction: by setting the shifting velocity to the largest convection velocity, error reduction should be obtained for all scales, albeit at different levels.

\begin{figure}
	\centering
	\includegraphics[width=0.9\linewidth]{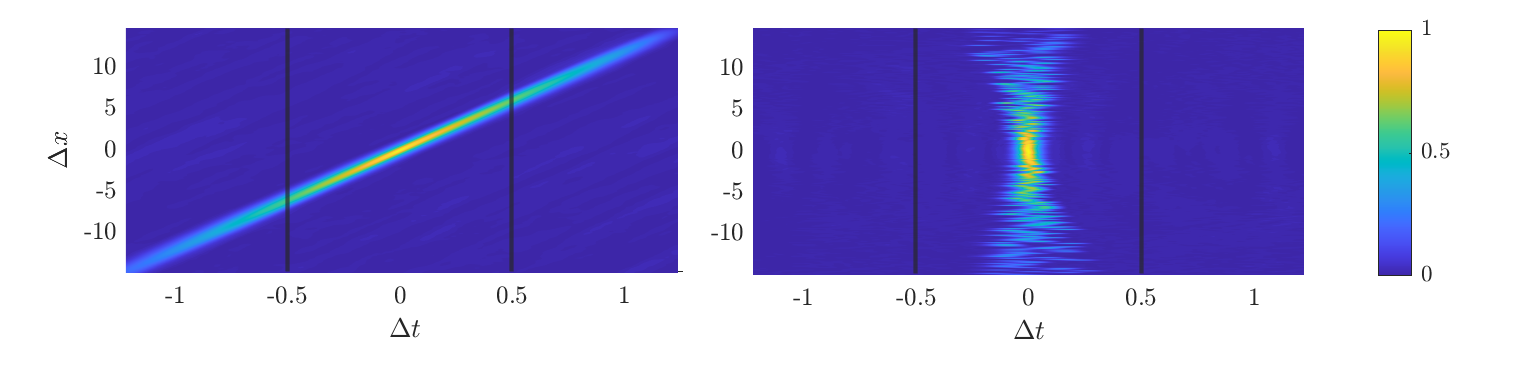}
	\caption{Space-time cross-correlation with respect to the point $x_p=15$ for case 2 ($U=12$, $A=1$). Vertical lines represent the bounds of a window of size $N_{FFT}=100$. Other cases display similar features. (Left) No shift; (Right) Random $U_{shift}(x) \in [0.8 \: U, 1.2 \: U]$.}
	\label{fig:GL_xcorr_random}
\end{figure}

Finally, we consider a random shift where $U_{shift}(x) \in [0.8 \: U, 1.2 \: U]$, whose space-time correlation map is displayed in figure \ref{fig:GL_xcorr_random}. Even though the shift is not constant nor optimal, it still efficiently reduces errors by maintaining the bulk of correlation within the window size bounds. With this setup we stress that the shifting velocity is not physical and not limited by conditions of continuity and smoothness, as phases in the Fourier space are corrected point by point, individually. Furthermore, we show that the exact match between shifting and local convection velocities is not a requirement, but rather a condition for optimal error reduction.

\subsubsection{Effect on different frequencies}

In a last analysis, we divert the focus from the frequency $\omega=2 \pi$ used in previous results to discuss the effect of the shifted SPOD on the ensemble of frequencies.

\begin{figure}
    \centering
    \includegraphics[width=\linewidth]{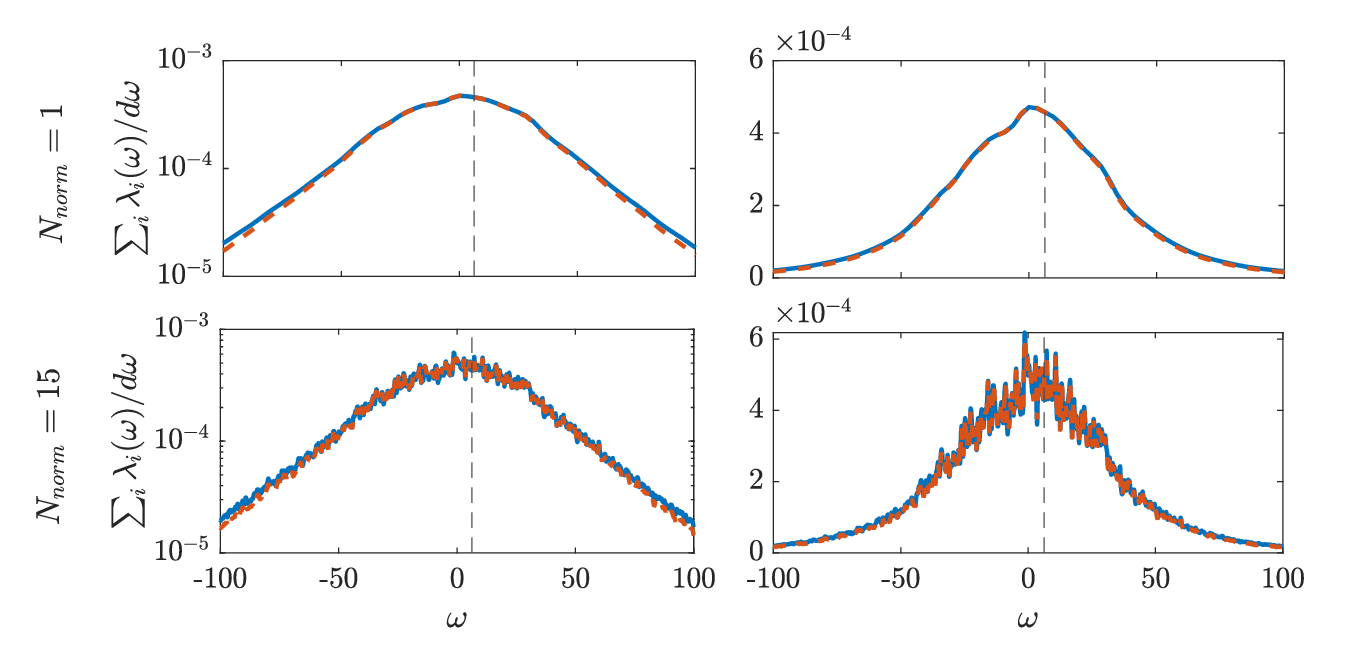}\\
    \caption{SPOD power spectra, for the case 3. Other cases display similar features.\\ Legend: ($\color{blue}\blacksquare$) Standard SPOD, ($\color{red}\blacksquare$) Shifted SPOD, ($- -$) $\omega=2 \pi$.}
    \label{fig:spectrum_GL}
\end{figure}  

Figure \ref{fig:spectrum_GL} illustrates SPOD power spectra computed from the sum of eigenvalues at each frequency for case 3. As we consider the complex-valued Ginzburg-Landau equation, the full spectrum of positive and negative frequencies is considered. At $N_{norm}=1$, the errors of modes at frequency $\omega=2 \pi$ for shifted and non-shifted methods are at least an order of magnitude apart. However, for both $N_{norm}=1$ and $N_{norm}=15$ the methods yield almost identical spectra, with discrepancies arising only for higher, less energetic frequencies.

\begin{figure}
    \centering
    \includegraphics[width=\linewidth]{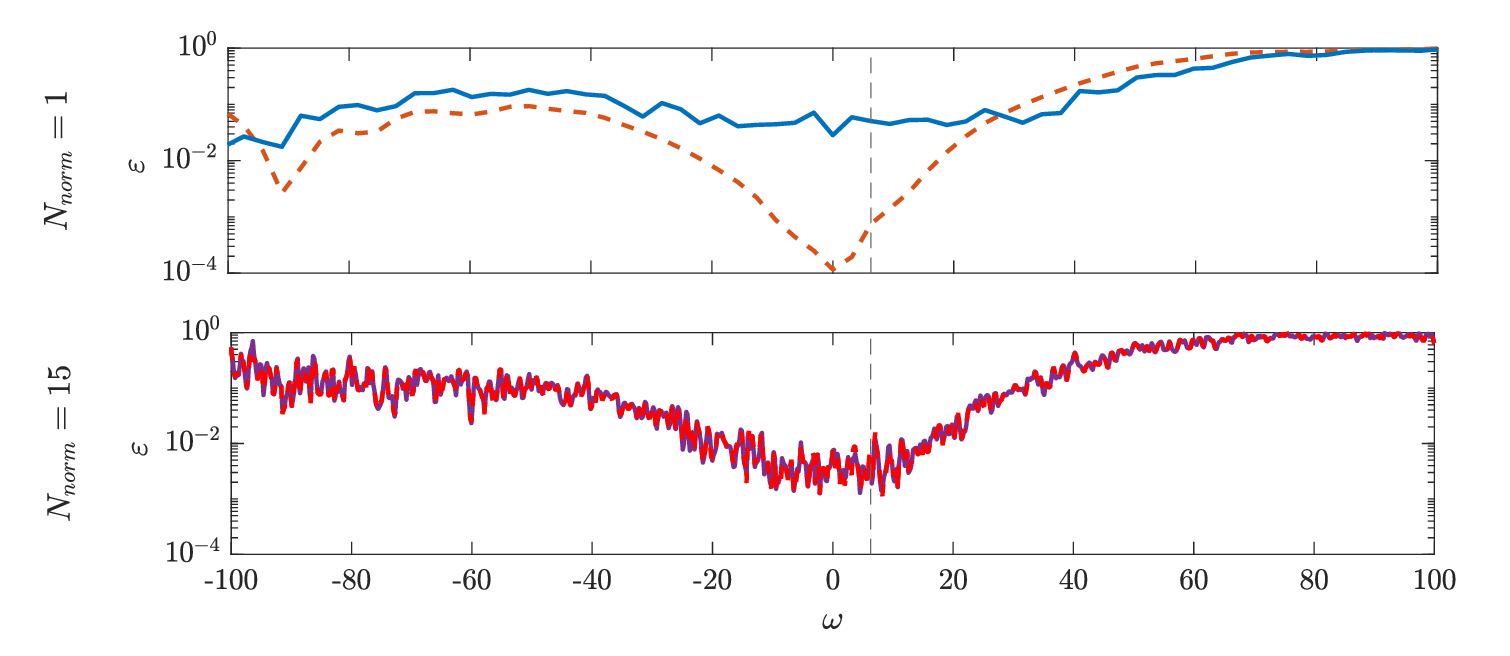}\\
    \caption{Estimation errors $\varepsilon$ as a function of frequency, for the case 3. Other cases display similar features. Legend: ($\color{blue}\blacksquare$) Standard SPOD, ($\color{red}\blacksquare$) Shifted SPOD, ($- -$) $\omega=2 \pi$.}
    \label{fig:error_freq_GL}
\end{figure}  

The behaviour of mode convergence can be further analysed in figure \ref{fig:error_freq_GL}, where estimation errors $\varepsilon$ are plotted with respect to the frequency. For a larger window size ($N_{norm}=15$), both shifted and standard algorithms converge to the same behaviour in frequency, corroborating the properties deduced in Appendix \ref{appB}. On the other hand, at $N_{norm}=1$, the most significant impact of the shifted SPOD is perceived close to the lowest frequencies, region where the energy separation is the greatest and where the standard SPOD achieves better convergence at $N_{norm}=15$. In a typical Navier-Stokes solution, these low frequencies correspond to the largest and most energetic scales.

\subsection{Transitional boundary layer over a flat plate}
\label{sec:simson}

In this section, the SPOD and resolvent methods are applied to study the most energetic/amplified structures present in a stable boundary layer subject transient growth prior to bypass transition. The case will serve as a numerical application of the concepts discussed in previous sections.

The database used for this analysis was generated using the SIMSON solver \cite{chevalier2007simson}. A large-eddy simulation (LES) of the transitional region of a Blasius-type boundary layer is carried out for the flow over a flat plate without leading edge and subject to no pressure gradient. This same database has already been validated and produced results in previous works \cite{sasaki_morra_cavalieri_hanifi_henningson_2020}.

\begin{figure}
	\centering
	\begin{minipage}{\textwidth}
		\centering
		\includegraphics[width=\linewidth]{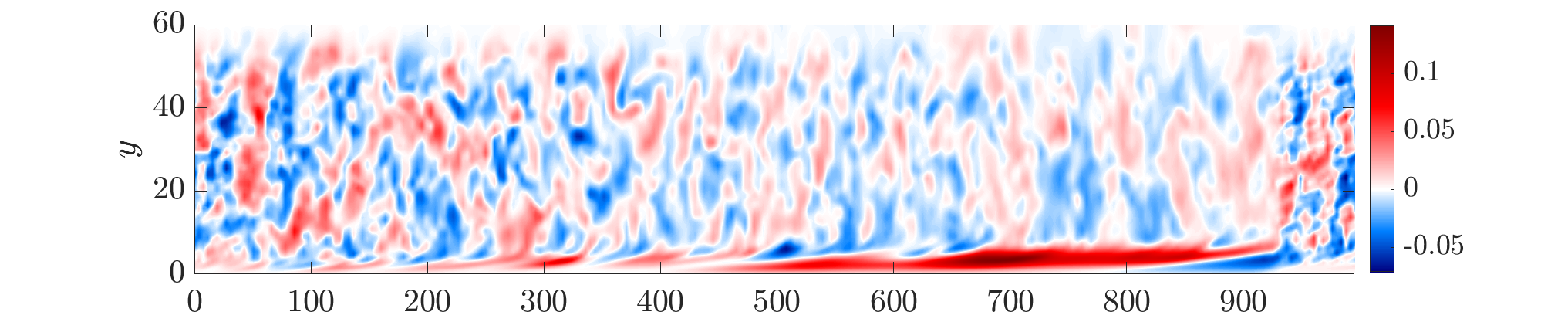}\\
	\end{minipage}\hfill
	\begin{minipage}{\textwidth}
		\centering
		\includegraphics[width=\linewidth]{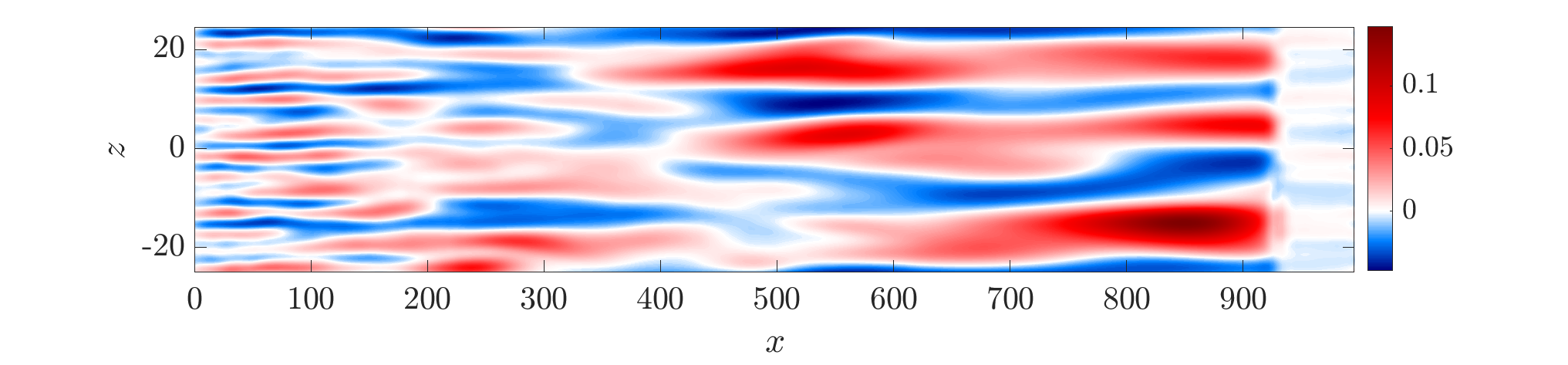}\\
	\end{minipage}
	\caption{Snapshot of the LES, where elongated structures can be observed inside the boundary layer. Colours represent stream-wise velocity component. Above, slice at $z=0$. Below, slice at $y=0.8$. The fringe zone is visible after $x=900$.}
	\label{fig:Tu30_snapshot}
\end{figure}

The computational domain (figure \ref{fig:Tu30_snapshot}) consists of a $231 \times 121 \times 108$ $(X \times Y \times Z)$ grid, composed by Chebyshev nodes in the direction perpendicular to the wall and homogeneously spaced points in the other two directions, which are discretised with Fourier modes. The $X$ axis points in the stream-wise direction and $x \in [0,1000]$, while the $Y$ axis is perpendicular to the wall and $y \in [0,60]$. The $Z$ axis follows the right hand rule and $z \in [-25,25]$. 

The simulation is periodic in the spam-wise direction, and periodicity along the stream-wise direction is assured by the introduction of a fringe region comprising positions $x_{fringe} \in [900,1000]$. All variables are non-dimensionalised by the free-stream velocity $U_\infty$ and displacement thickness $\delta^*_0$ at the intake, position where $Re^*= U_\infty \delta^*_0/\nu = 300$, with $\nu$ being the kinematic viscosity of the fluid.

In order to produce isotropic turbulence at the free stream, a number of modes from the continuous branch of the Orr-Sommerfeld and Squire operators are forced in the fringe region, with turbulence intensity of $3.0\%$ measured in $rms$ level, in the same way of \cite{sasaki_morra_cavalieri_hanifi_henningson_2020}, following \cite{brandt_schlatter_henningson_2004}. The resulting database is composed of 2000 snapshots of fully developed, statistically stationary velocity fluctuations fields around the Blasius base flow, computed with a constant time step of $\delta t = 10$. The domain is long enough to display the growth of streaks in the laminar upstream region, but ends before the development of turbulent spots, as seen in figure \ref{fig:Tu30_snapshot}.

\subsubsection{SPOD and shifted SPOD}
\label{sec:SPOD_BL}

\begin{figure}
	\centering
	\begin{minipage}{\textwidth}
		\centering
		\includegraphics[width=\linewidth]{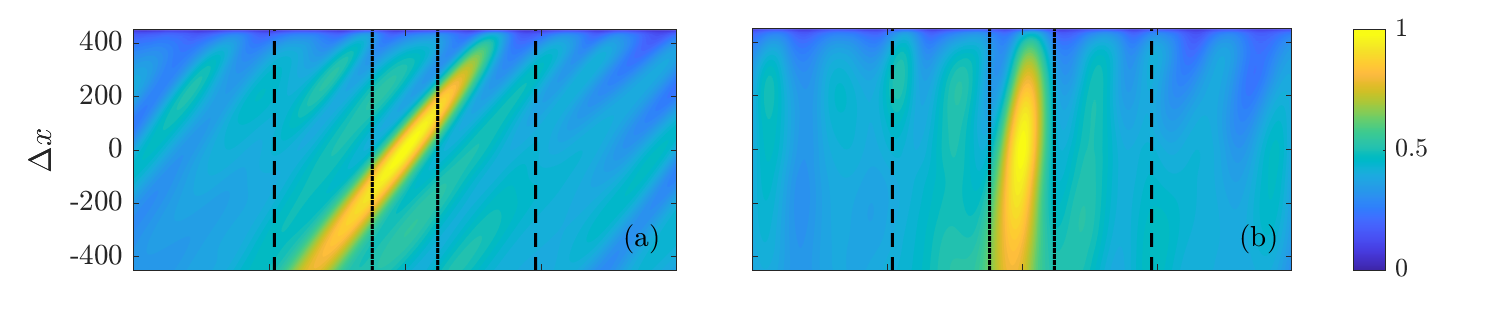}\\
	\end{minipage}\hfill
	\begin{minipage}{\textwidth}
		\centering
		\includegraphics[width=\linewidth]{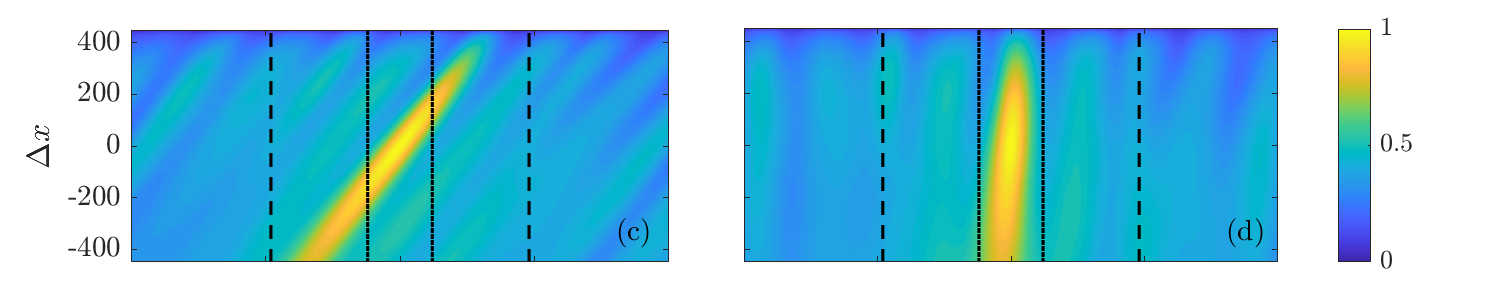}\\
	\end{minipage}
	\begin{minipage}{\textwidth}
		\centering
		\includegraphics[width=\linewidth]{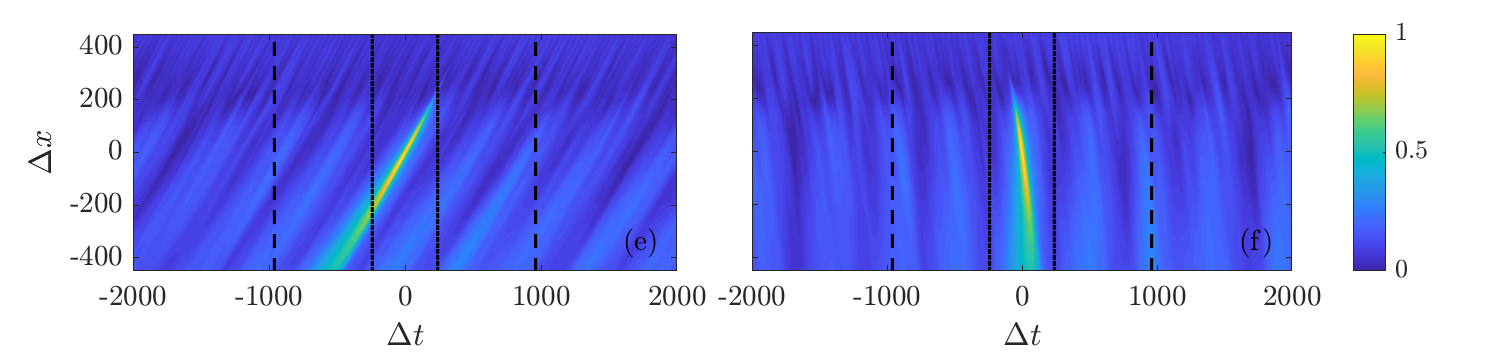}\\
	\end{minipage}
	\caption{Space-time cross-correlation for $\beta=0.377$ of the $u$ velocity component at different distances of the wall. (a,b) $y = 0.01$; (c,d) $y=1$; (e,f) $y=6$; (a,c,d) Non-shifted field; (b,d,f) Shifted field. (Dashed line) $N_{FFT} = 192$. (Dotted line) $N_{FFT} = 48$.}
	\label{fig:BL_xcorr}
\end{figure}  

Since the domain is periodic in $Z$, a Fourier decomposition can be performed in this direction and each wave number $\beta$ can be analysed separately, making the velocity field effectively 2D for each $\beta$. Following the definitions of section \ref{sec:SPOD}, the state vector $\mathbf{q}$ is defined according to eq. (\ref{eq:state}) and the inner-product weight matrix $\mathbf{W}$ is designed to account for the grid quadrature. SPOD modes are computed for $x \in [0,900]$, excluding the fringe zone. We employ the same windowing function from equation (\ref{eq:wind_func}) which allows an overlap of 75\% between blocks.

In order to better study the effects of window sizing, we consider two cases with a time series division in $N_{FFT_1}=48$ and $N_{FFT_2}=192$ snapshots per block, leading to, respectively, $N_{b_1}=330$ and $N_{b_2}=38$ blocks with the standard method and $N_{b_1}=310$ and $N_{b_2}=36$ with the shifted one. As the streaky structures involved in the bypass transition have very low characteristic frequency and slow dynamics \cite{nogueira_cavalieri_jordan_jaunet_2019,pickering_rigas_nogueira_cavalieri_schmidt_colonius_2020,nidhan2020}, we focus the analysis on SPOD modes at the lowest non-zero frequency in case 1, $\omega=0.0131$.

Following the steps described in section \ref{sec:choice}, for the shifted SPOD algorithm, (i) we choose to compute time lags with respect to the stream-wise position $x_p=450$, the farthest possible from the influence of the fringe region. Next, in figure \ref{fig:BL_xcorr}(a,c,e), (ii) we plot the space-time correlations for the stream-wise component $u$ at different distances from the wall. (iii) A constant shifting velocity $U_k=0.75$ (eq. \ref{eq:shift}) was fixed for all points. In this case, the shifting velocity does not match the free stream velocity $U_\infty=1$. This was carefully set to align approximately the cross-correlations of points inside the boundary layer, as shown in figure \ref{fig:BL_xcorr}(b,d,f). (iv) The alignment of peaks generated by the shift allows the bulk of correlation, that required a window of size $N_{FFT}=192$, to be captured in a much smaller window of $N_{FFT}=48$.

\begin{figure}
	\centering
	Case 1: $N_{FFT} = 48$\\
	\hspace{0.2cm}
	\begin{minipage}{\textwidth}
		\centering
		(a) Normalised energy\\
		\includegraphics[width=0.8\linewidth]{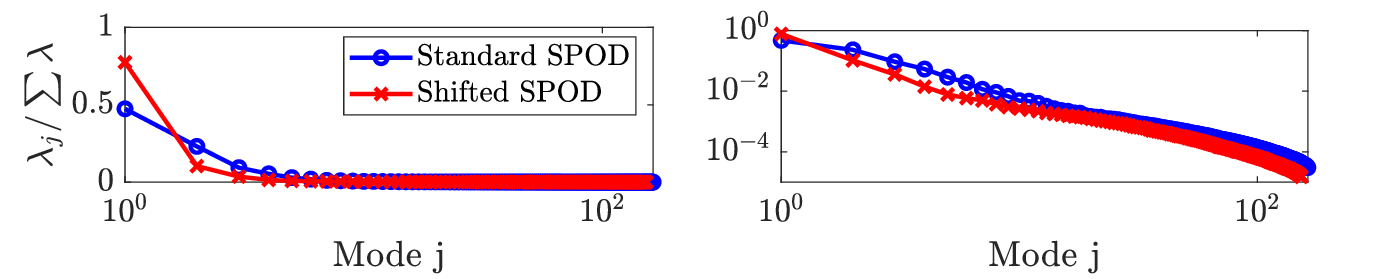}
	\end{minipage}\\
	\begin{minipage}{0.5\textwidth}
		\centering	
		\includegraphics[width=0.95\linewidth]{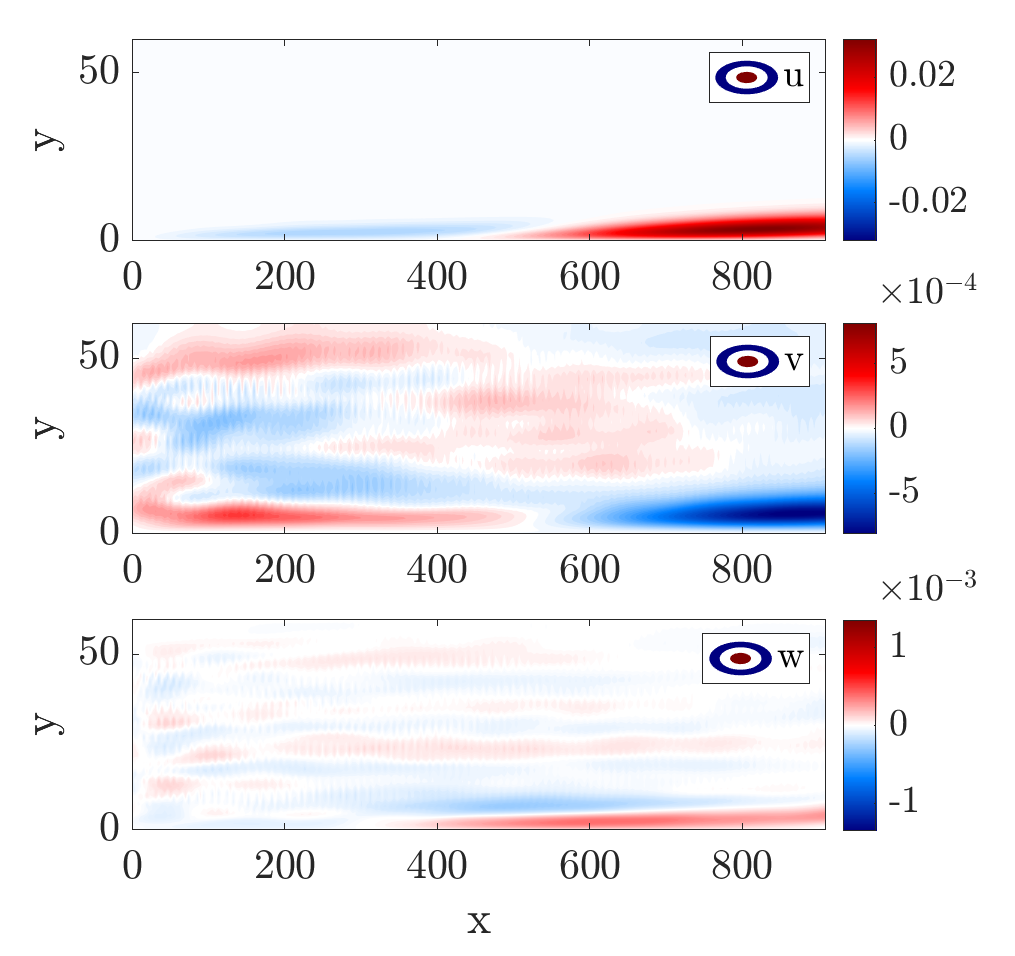}\\
		(b) Standard SPOD
	\end{minipage}\hfill
	\begin{minipage}{0.5\textwidth}
		\centering
		\includegraphics[width=0.95\linewidth]{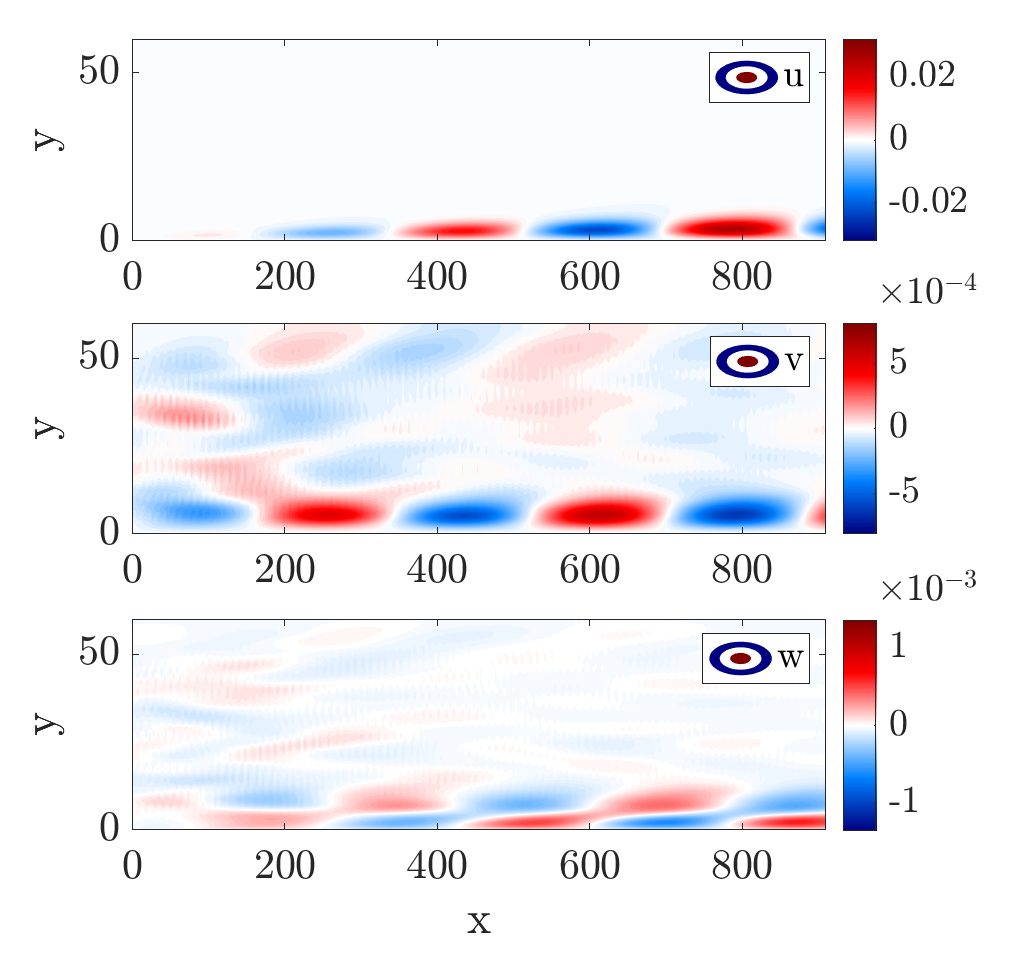}\\
		(c) Shifted SPOD
	\end{minipage}
	\rule{0.9\textwidth}{0.5pt}\\
	Case 2: $N_{FFT} = 192$\\
	\hspace{0.2cm}
	\begin{minipage}{\textwidth}
		\centering
		(d) Normalised energy\\
		\includegraphics[width=0.8\linewidth]{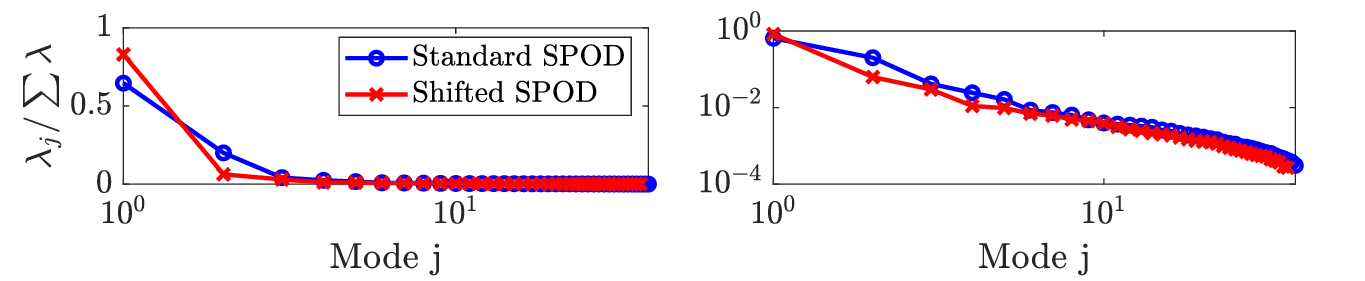}
	\end{minipage}\\
	\begin{minipage}{0.5\textwidth}
		\centering
		\includegraphics[width=0.95\linewidth]{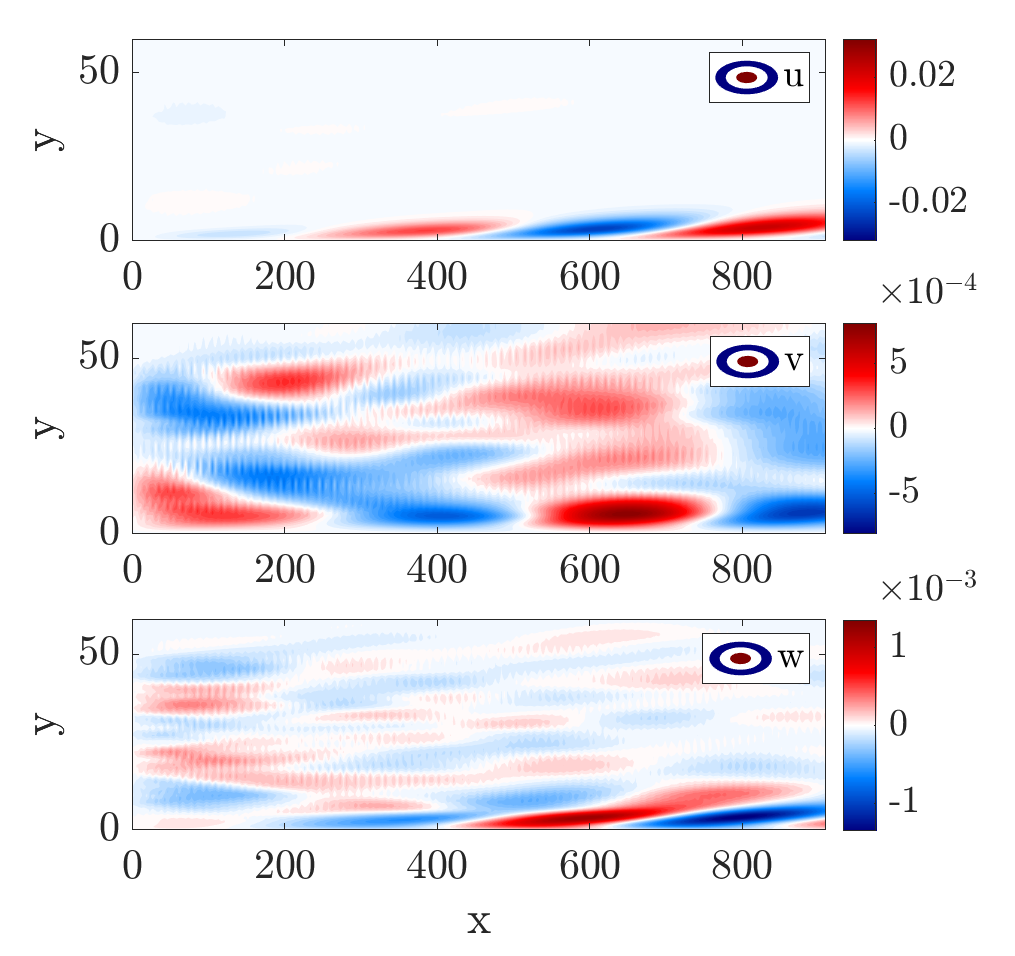}\\	
		(e) Standard SPOD
	\end{minipage}\hfill
	\begin{minipage}{0.5\textwidth}
		\centering
		\includegraphics[width=0.95\linewidth]{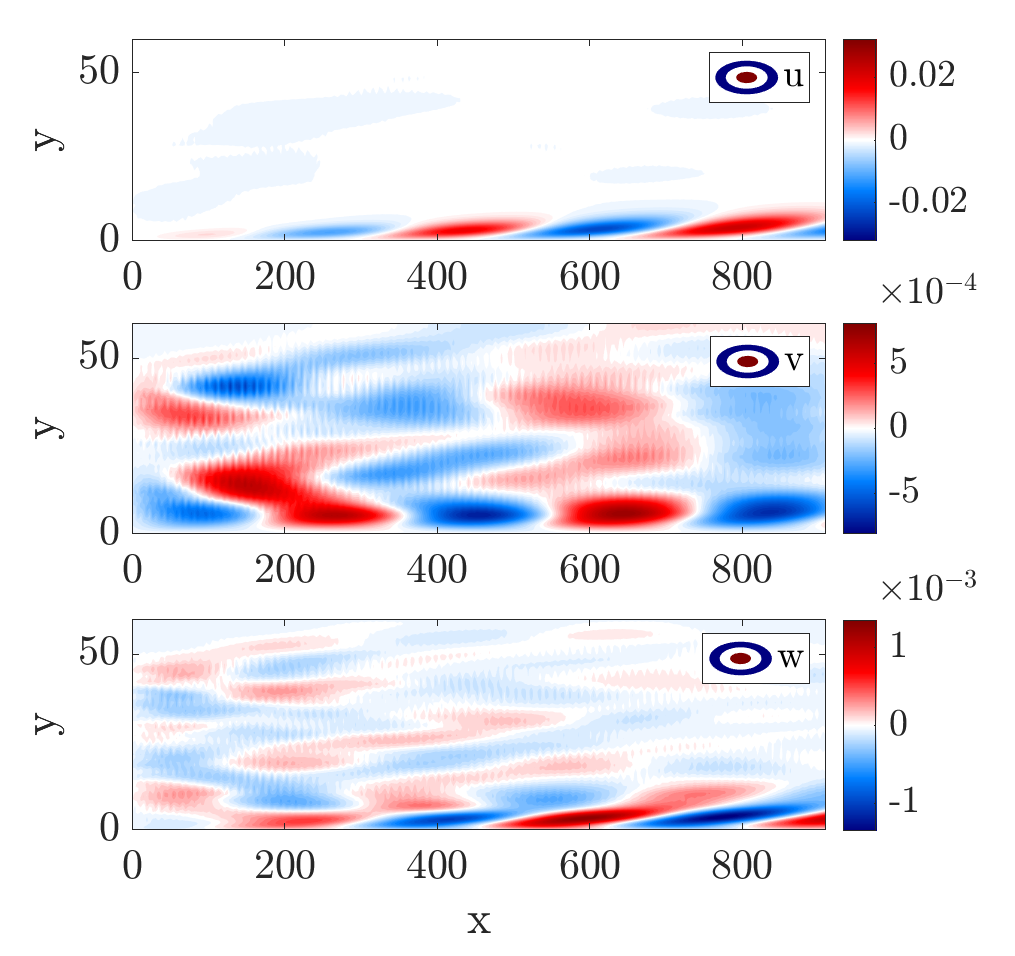}\\
		(f) Shifted SPOD
	\end{minipage}
	\caption{Results using a window of size $N_{FFT}=48$ and $N_{FFT}=192$ at $\beta = 0.377$, $\omega=0.0131$. (a,d) Eigenvalues normalised by total energy. (b,c,e,f) Real part of the leading mode.}
	\label{fig:mode1_omega0.0131}
\end{figure}

Results for the most energetic spam-wise wave number $\beta = 0.377$ at frequency $\omega=0.0131$ are displayed in figure \ref{fig:mode1_omega0.0131}. Figures \ref{fig:mode1_omega0.0131}(a,d) display eigenvalues as a fraction of total energy. With both methods, the first mode dominates the flow dynamics. At $N_{FFT}=48$, the first standard SPOD mode accounts for $47.3\%$ of the energy, while the corresponding shifted SPOD mode captures $77.4\%$. At $N_{FFT}=192$, these values correspond to $64.6\%$ and $82.9\%$ respectively. In both cases, the shifted SPOD is able to capture more energy in the first mode.

We can quantify the similarity between modes computed with different windows sizes by calculating alignment coefficients 
\begin{equation}\label{eq:alignment}
    \mu_{i,j}(\omega) = \frac{| \langle \mathbf{\hat{\psi}}_i(\omega),\mathbf{\hat{\psi}}_j(\omega) \rangle|}{||\mathbf{\hat{\psi}}_i(\omega)|| \cdot ||\mathbf{\hat{\psi}}_j(\omega)||} 
\end{equation} 
from which we get $\mu_{48,192} = 0.23$ for standard SPOD and $\mu_{48,192} = 0.88$ for shifted SPOD. The higher alignment between modes computed with the shifted algorithm puts in evidence its lower sensitivity with respect to the window size parameter.
\subsubsection{Comparison with resolvent analysis} 
\label{sec:global_resolvent}

With the purpose of providing a deeper insight of the intrinsic dynamics of the boundary layer system, global resolvent modes, computed according to the method presented in \cite{kaplan_jordan_cavalieri_bres_2021} and \cite{abreu_tanarro_cavalieri_schlatter_vinuesa_hanifi_henningson_2021}, are compared with the available SPOD modes. The governing Navier-Stokes equations are linearised around a Blasius profile base flow and cast in an input-output form as described in section \ref{sec:resolvent}. Even though this is a non-linear system, it is presumed that, when the resolvent gains display an expressive separation in magnitude, the first SPOD and resolvent response modes are similar regardless of the non-linear forcing statistics \cite{beneddine_sipp_arnault_dandois_lesshafft_2016}. This effect has already been observed in turbulent flows \cite{schmidt_towne_rigas_colonius_bres_2018,10.1115/1.4042736,PhysRevFluids.4.063901,abreu_cavalieri_schlatter_vinuesa_henningson_2020}.

\begin{figure}
	\centering
	\includegraphics[width=\linewidth]{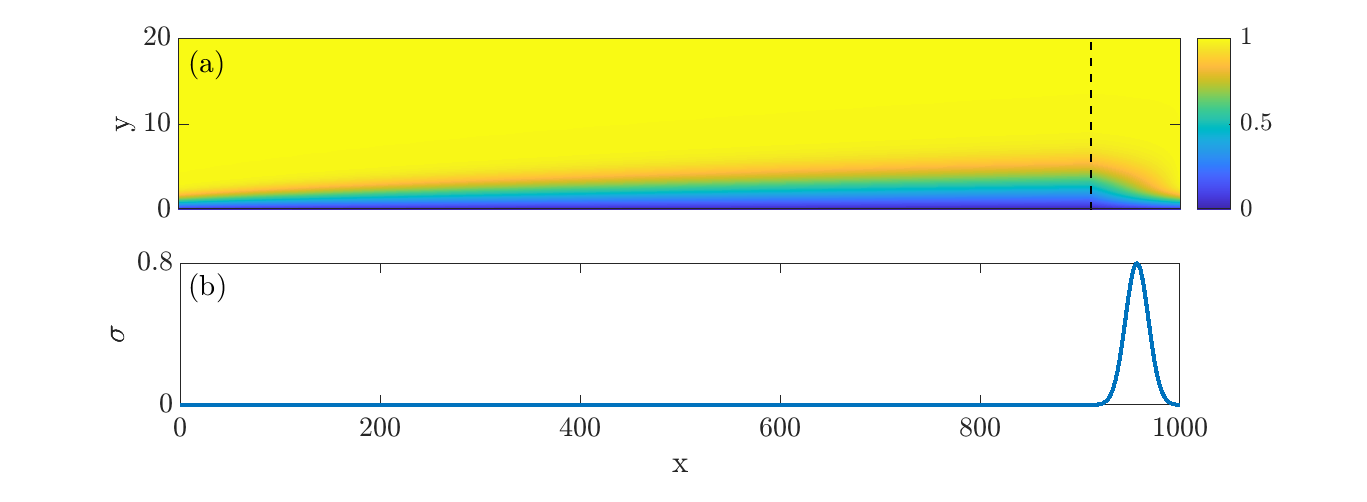}\\
	\caption{(a) Blasius mean velocity field ($U$ component). The dashed line indicates the beginning of the fringe region. (b) Damping factor. The maximum of $\sigma$ matches the value used in the LES.}
	\label{fig:damping_fringe}
\end{figure}

The global resolvent computational domain was defined similarly to the LES, with the same domain size discretised with a $512 \times 121$ $(X \times Y)$ grid. We introduce a fringe zone in $x_{fringe} \in [900,1000]$ where fluctuations are damped and the base flow field is artificially treated to enforce the periodicity in X (figure \ref{fig:damping_fringe}). Homogeneous boundary conditions for fluctuations are set at the lower and upper domain limits. Derivatives are approximated by a fourth-order finite differences scheme in $X$ and Chebyshev polynomials in $Y$.

As in the LES, all variables are non-dimensionalised by the displacement thickness $\delta^*$ and free-stream velocity $U_\infty = 1$ at the intake, where $Re^*= 300$. Periodicity in the span and stationarity imply the normal mode ansatz
\begin{equation}\label{eq:normal_mode}
	\mathbf{\tilde{q}}(x, y, z, t) = \mathbf{\hat{q}}(x,y) \exp[i(\beta z-\omega t)],
\end{equation}
where the state vector $\mathbf{q} = \left[u,v,w,p\right]^T$ contains the velocity components and the pressure stacked. The weight matrices $\mathbf{W}_y$ and $\mathbf{W}_f$ were set to account for the domain quadrature and the operator $\mathbf{B}$ restricts the forcing application to the space outside the fringe zone. The observation operator $\mathbf{H}$  removes the pressure component from the output, as the flow is incompressible and only the velocity field is analysed. Even though the state $\mathbf{q}$ contains point in the fringe, the resolvent operator gains only consider points outside of it,  by a proper definition of $\mathbf{H}$.

\begin{figure}
	\centering
	(a)\\
	\begin{minipage}{0.5\textwidth}
		\centering
		\includegraphics[width=0.8\linewidth]{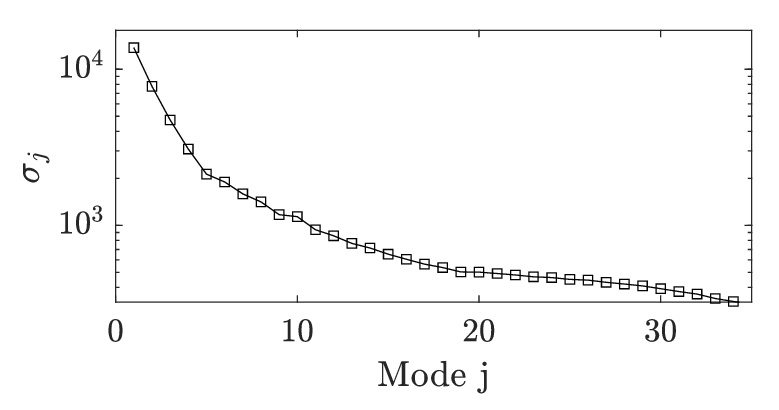}\\	
	\end{minipage}\hfill\\
	\begin{minipage}{0.5\textwidth}
		\centering
		\includegraphics[width=0.95\linewidth]{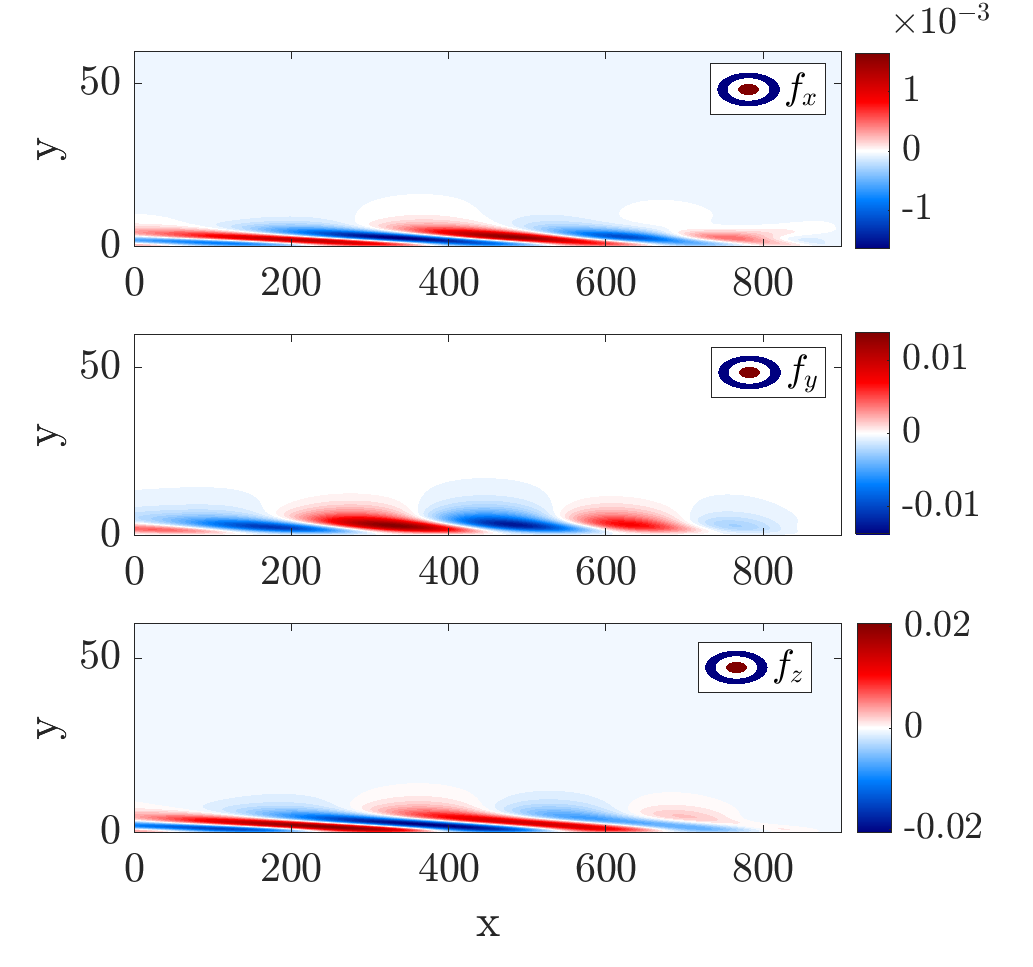}\\
		(b)
	\end{minipage}\hfill
	\begin{minipage}{0.5\textwidth}
		\centering
		\includegraphics[width=0.95\linewidth]{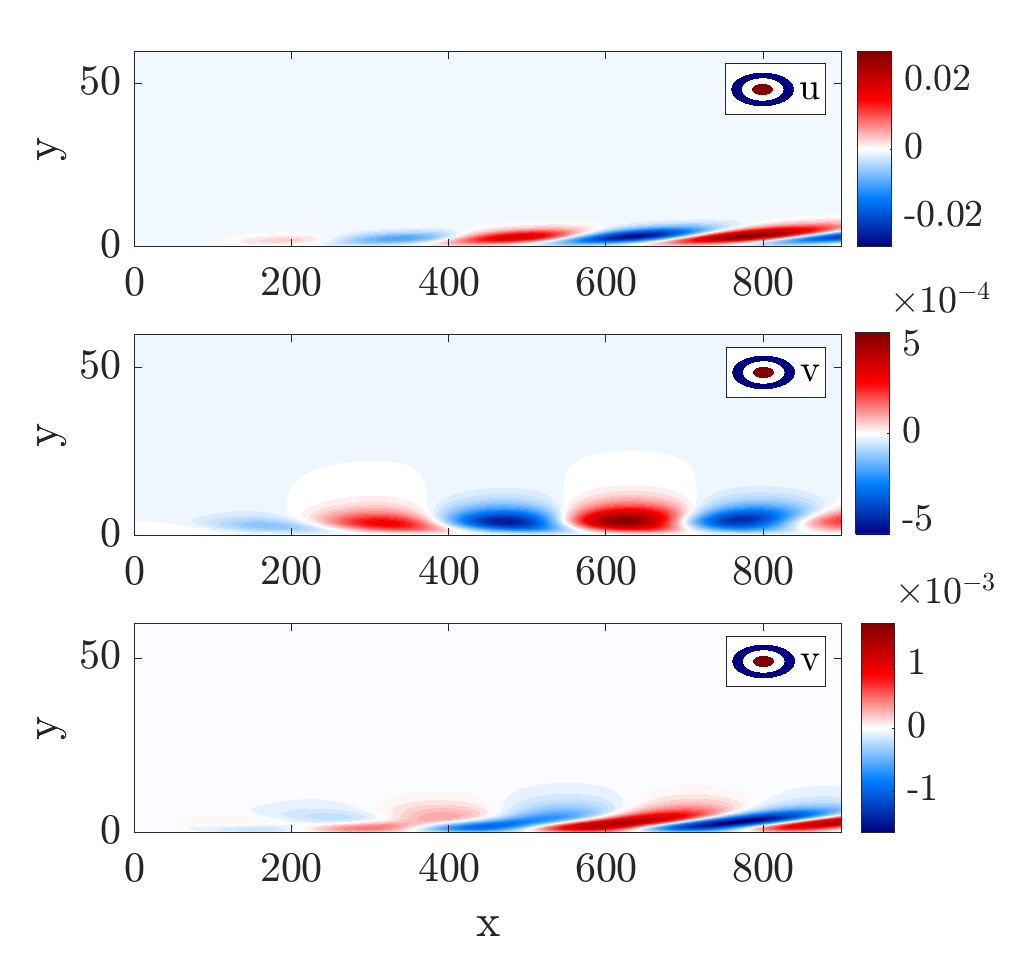}\\
		(c)
	\end{minipage}\hfill
	\caption{(a) Resolvent gains. (b) Real part of leading forcing mode. (c) Real part of leading response mode.}
	\label{fig:resolvent_mode1_omega0.0131}
\end{figure}

\begin{figure}
    \centering
    \includegraphics[width=\linewidth]{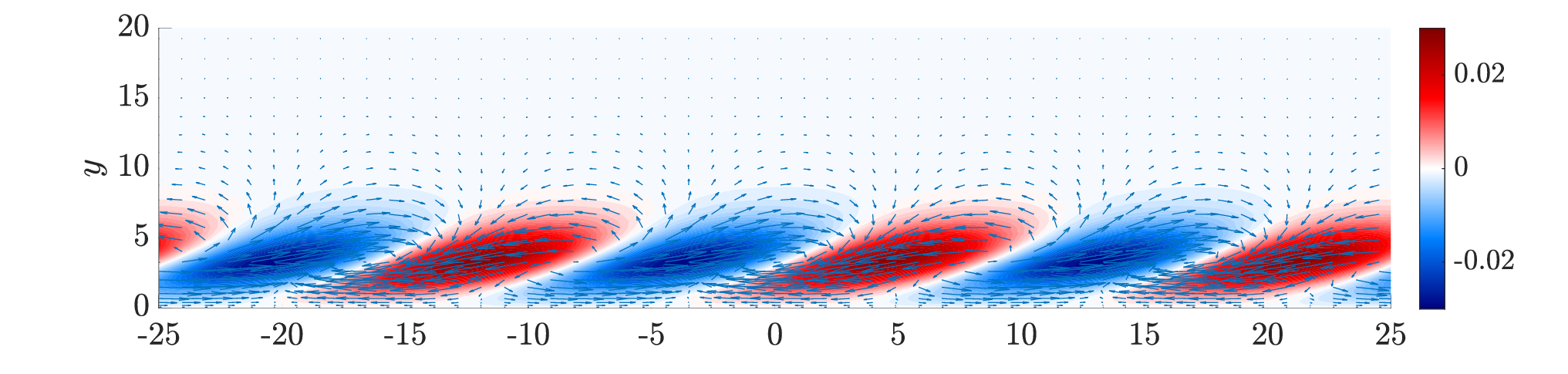}\\	
    (a) Resolvent response\\
    \rule{0.9\textwidth}{0.5pt}\\
    Case 1: $N_{FFT} = 48$\\
    \includegraphics[width=\linewidth]{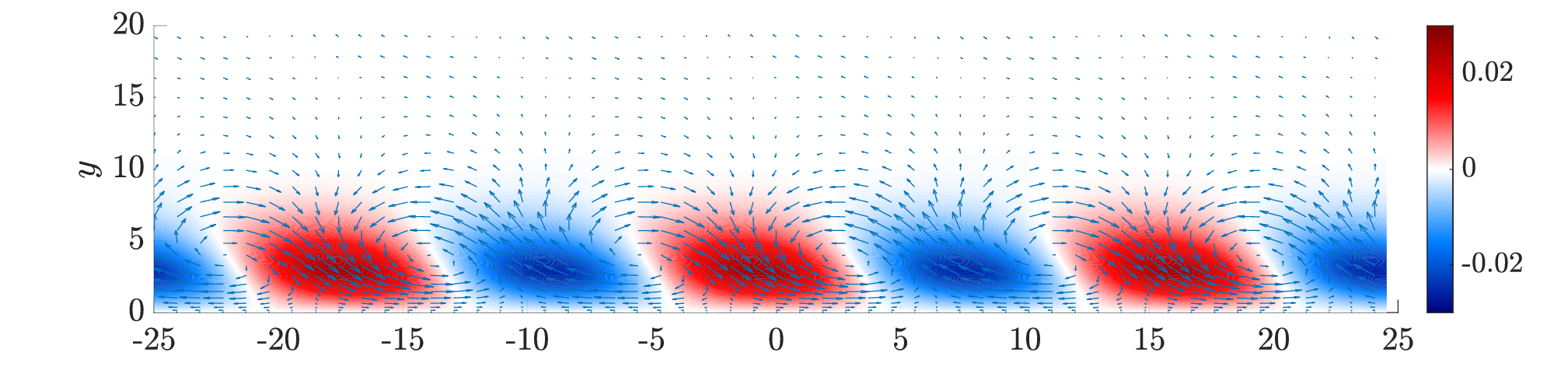}\\
    (b) Standard SPOD\\
    \includegraphics[width=\linewidth]{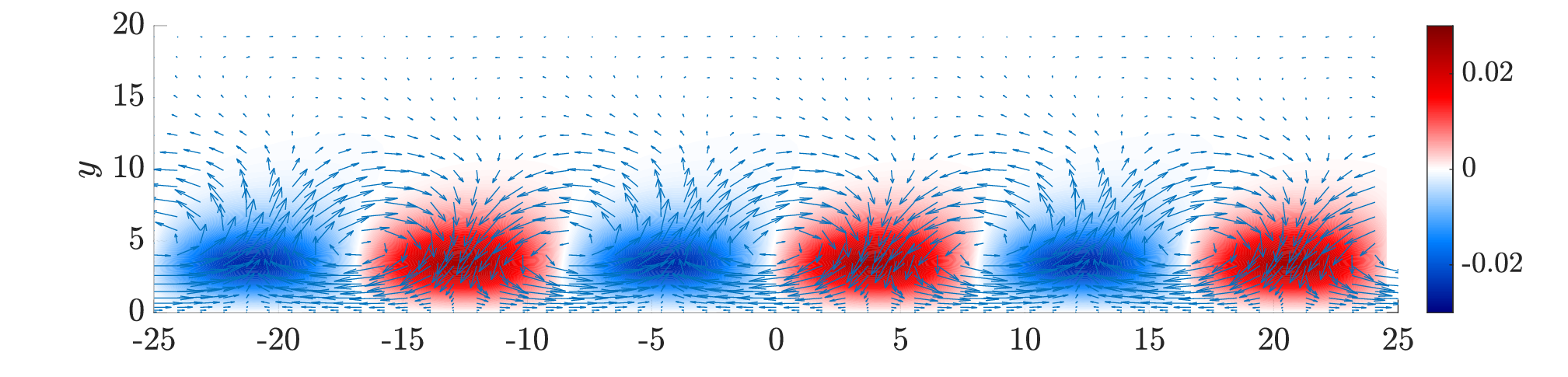}\\
    (c) Shifted SPOD\\
    \rule{0.9\textwidth}{0.5pt}\\
    Case 2: $N_{FFT} = 192$\\
    \includegraphics[width=\linewidth]{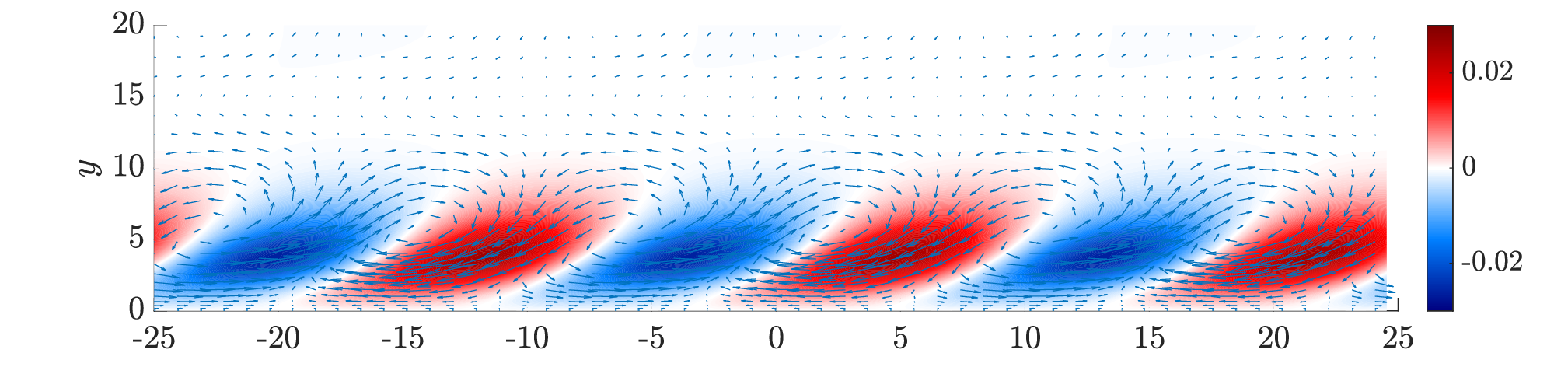}\\
    (d) Standard SPOD\\
    \includegraphics[width=\linewidth]{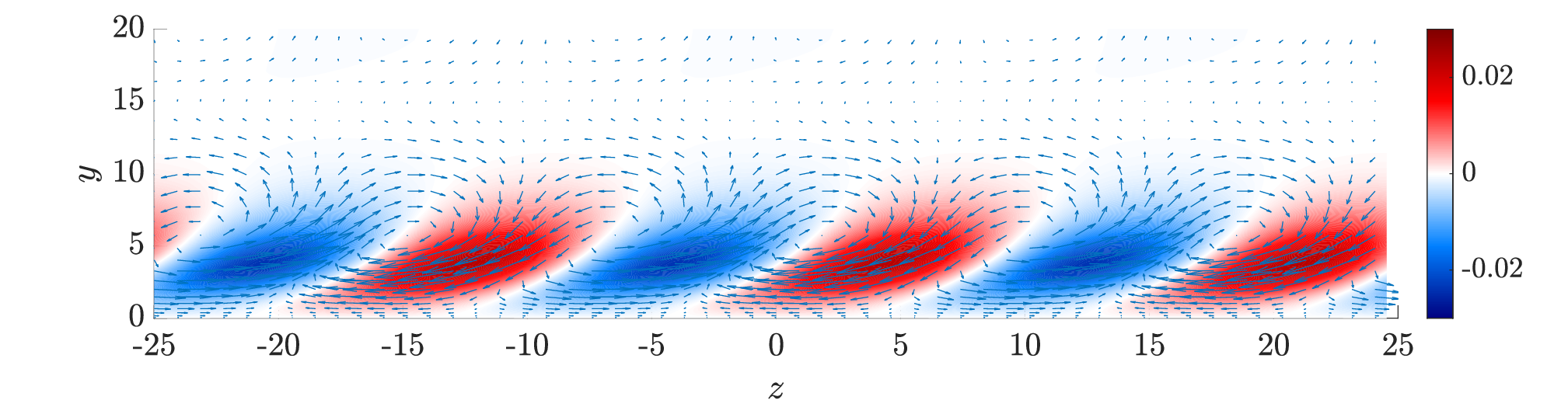}\\
    (e) Shifted SPOD\\
    \caption{Section $x=700$ of the reconstructed modes. Legend: (colours) $u$ velocity component; ($\rightarrow$) Perpendicular velocity component.}
    \label{fig:quiver_omega0.0131}
\end{figure}

\begin{figure}
    \centering
    \includegraphics[width=\linewidth]{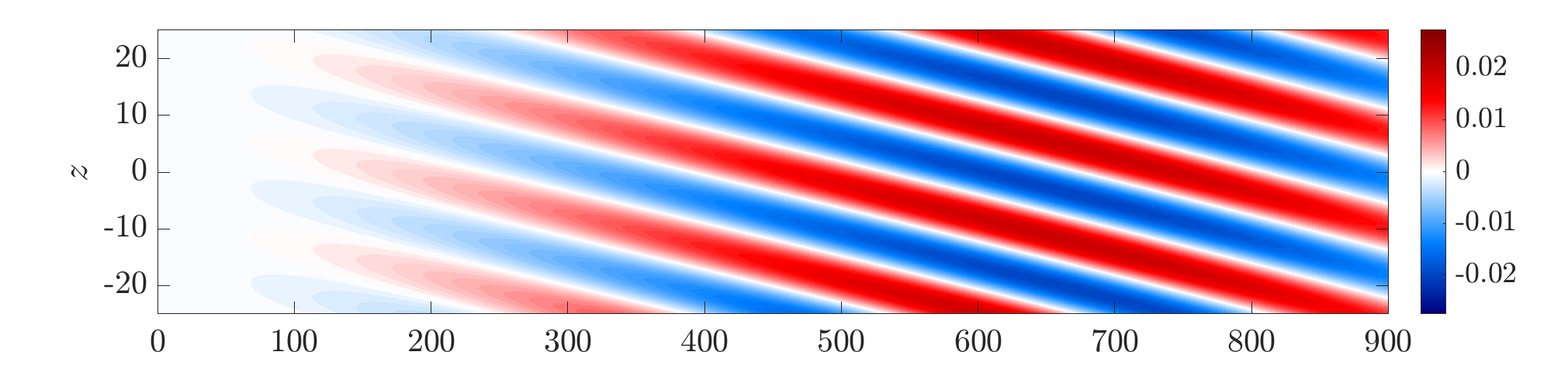}\\	
    (a) Resolvent response\\
    \rule{0.9\textwidth}{0.5pt}\\
    Case 1: $N_{FFT} = 48$\\
    \includegraphics[width=\linewidth]{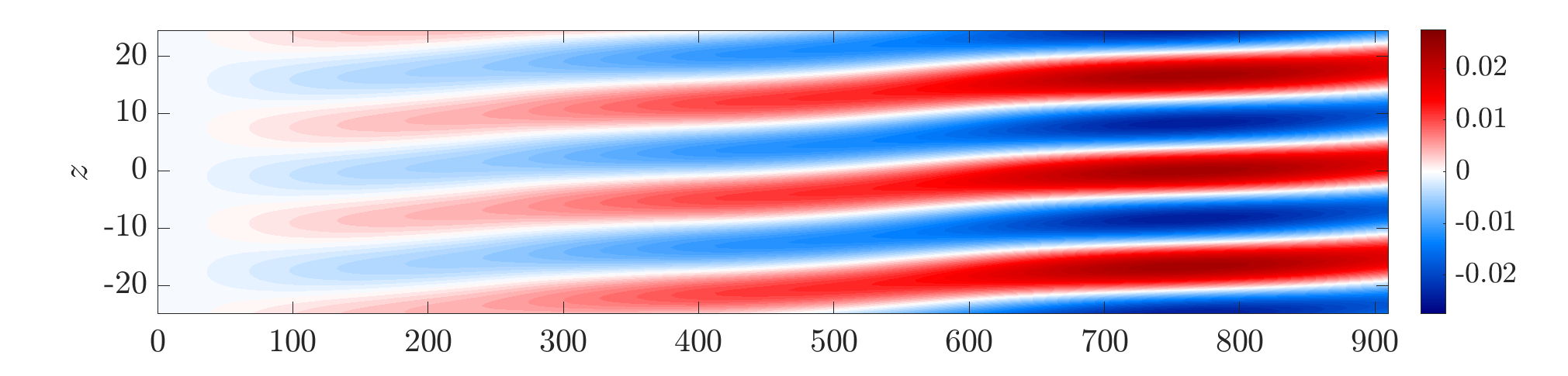}\\
    (b) Standard SPOD\\
    \includegraphics[width=\linewidth]{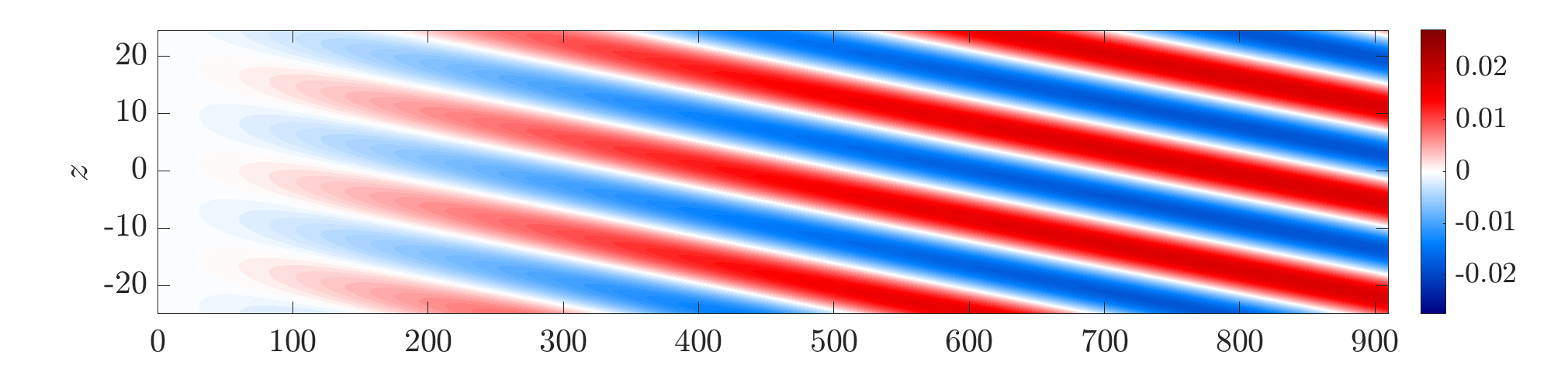}\\
    (c) Shifted SPOD\\
    \rule{0.9\textwidth}{0.5pt}\\
    Case 2: $N_{FFT} = 192$\\
    \includegraphics[width=\linewidth]{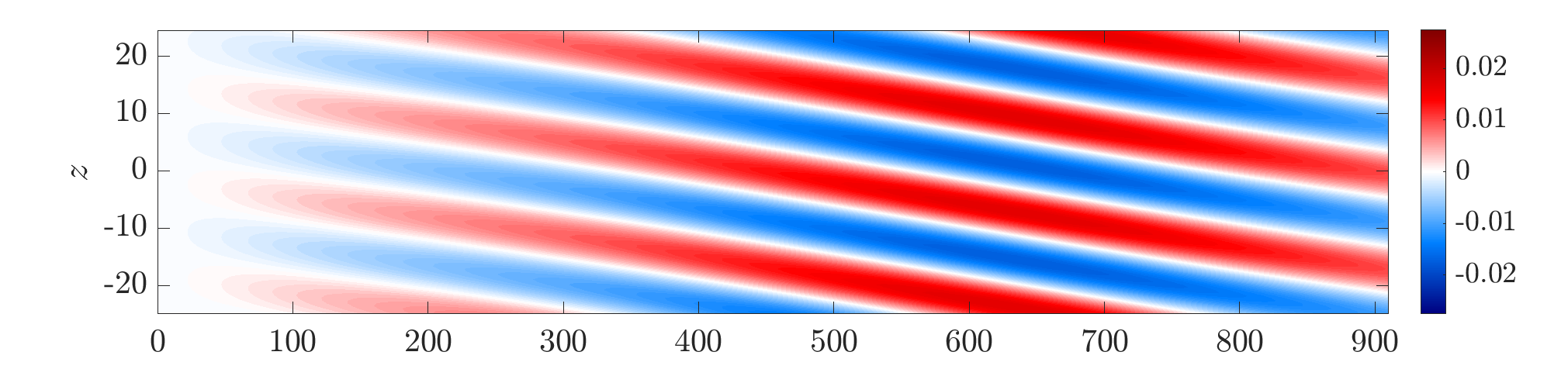}\\
    (d) Standard SPOD\\
    \includegraphics[width=\linewidth]{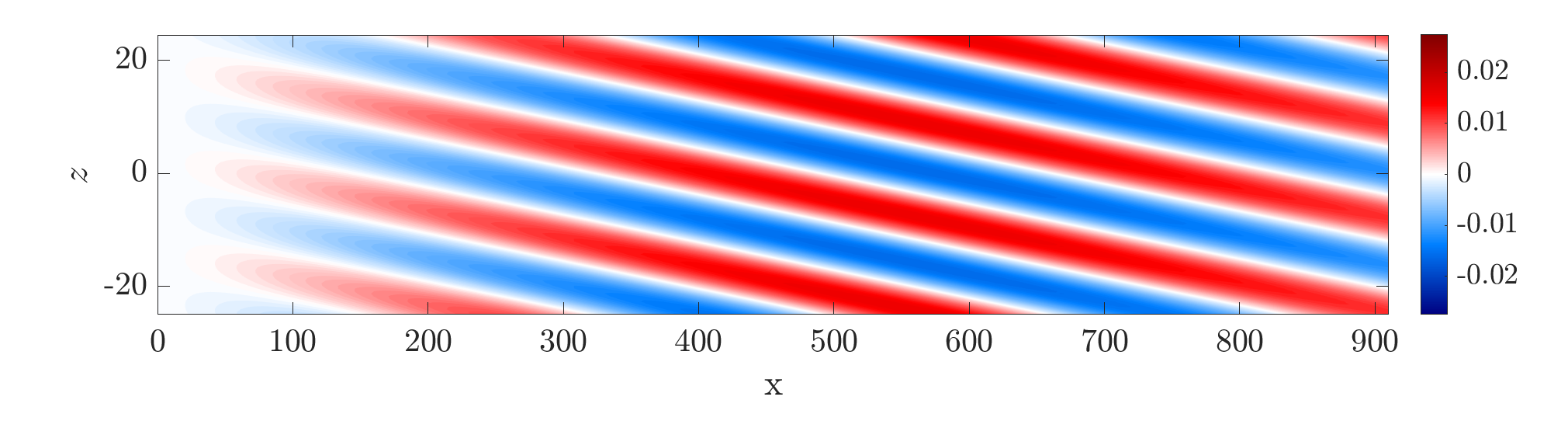}\\
    (e) Shifted SPOD\\
    \caption{Section $y=2$ of the reconstructed modes, inside the boundary layer. Legend: (colours) $u$ velocity component.}
    \label{fig:top_omega0.0131}
\end{figure}

Due to the significant size of matrices involved, the eigenproblem
\begin{equation}\label{eq:eig_globalresolvent}
	\begin{aligned}
	\mathbf{R}^H \mathbf{R} \mathbf{V} = \mathbf{V} \Sigma^2\\
	\mathbf{U} = \mathbf{R}	\mathbf{V} \Sigma^{-1}	
	\end{aligned}
\end{equation}
that calculates resolvent modes and gains is solved using sparse matrices, via a LU decomposition and 50 iterations of the Arnoldi method \cite{martini_rodriguez_towne_cavalieri_2021}, without explicitly inverting $\mathbf{L}$. The details of this method are described in \cite{kaplan_jordan_cavalieri_bres_2021}. Results of the global resolvent analysis for the same $\beta = 0.377$ and $\omega=0.0131$ as in the LES are displayed in figure \ref{fig:resolvent_mode1_omega0.0131}. Figures \ref{fig:quiver_omega0.0131} and \ref{fig:top_omega0.0131} show the comparison between the reconstructed 3D resolvent mode and the previously studied SPOD modes. 

Figure \ref{fig:resolvent_mode1_omega0.0131}a shows that the leading gain is sufficiently separated from the rest and thus the corresponding mode can be used as comparison with the SPOD. Besides that, we observe in figures \ref{fig:resolvent_mode1_omega0.0131}b and \ref{fig:resolvent_mode1_omega0.0131}c that the leading response and forcing resolvent modes demonstrate the mechanism described in the lift-up effect: a forcing perpendicular to the stream-wise direction, with small $f_x$ forcing component and larger $f_y$ and $f_z$, generating a growing response primarily in the $u$ velocity component. Such results are consistent with the earlier calculations by \cite{monokrousos_jfm_2010}. From figure \ref{fig:quiver_omega0.0131} it is clear that this dynamic generates intercalating regions of alternating stream-wise velocity and counter rotating stream-wise vortices. When the vorticity transports matter downwards to the boundary layer, a positive streak is created, while the opposite occurs when matter is forced upwards. 

For case 2, cross-sections of shifted and non-shifted SPOD modes are almost identical and match the dynamics of the most amplified linearised response. However, the same is not true for case 1, suggesting that streak skewness is an attribute sensitive to window size.

Other discrepancies are clear in figure \ref{fig:top_omega0.0131}. While the resolvent response and shifted modes feature aligned elongated structures, displaying the characteristic deviation caused by the phase velocity $\omega/\beta$, the non-shifted mode in case 1 has an opposite trend. Since phase velocity is not affected by the windowing, we conclude that this is a convergence issue of the standard SPOD, corrected by the proposed temporal shift operation. 

A growth of longitudinal vortices along the stream-wise direction is perceived in the SPOD modes (magnitudes of $v$ and $w$ components in figure \ref{fig:mode1_omega0.0131}). This feature is predicted by the resolvent analysis, as seen in figure \ref{fig:resolvent_mode1_omega0.0131} and constitutes a different behaviour from that caused by the transient growth, where vortices decay in $x$ while streaks grow \cite{doi:10.1063/1.869908,luchini2000}. These results imply that the free-stream turbulence provides a continuous forcing to the stream-wise vortices, which may be understood as projecting onto the forcing resolvent mode of figure \ref{fig:resolvent_mode1_omega0.0131}b and leading to stream-wise vortices and streaks that grown in $x$.

\begin{figure}
	\centering
	\begin{minipage}{0.5\textwidth}
		\centering
		\includegraphics[width=\linewidth]{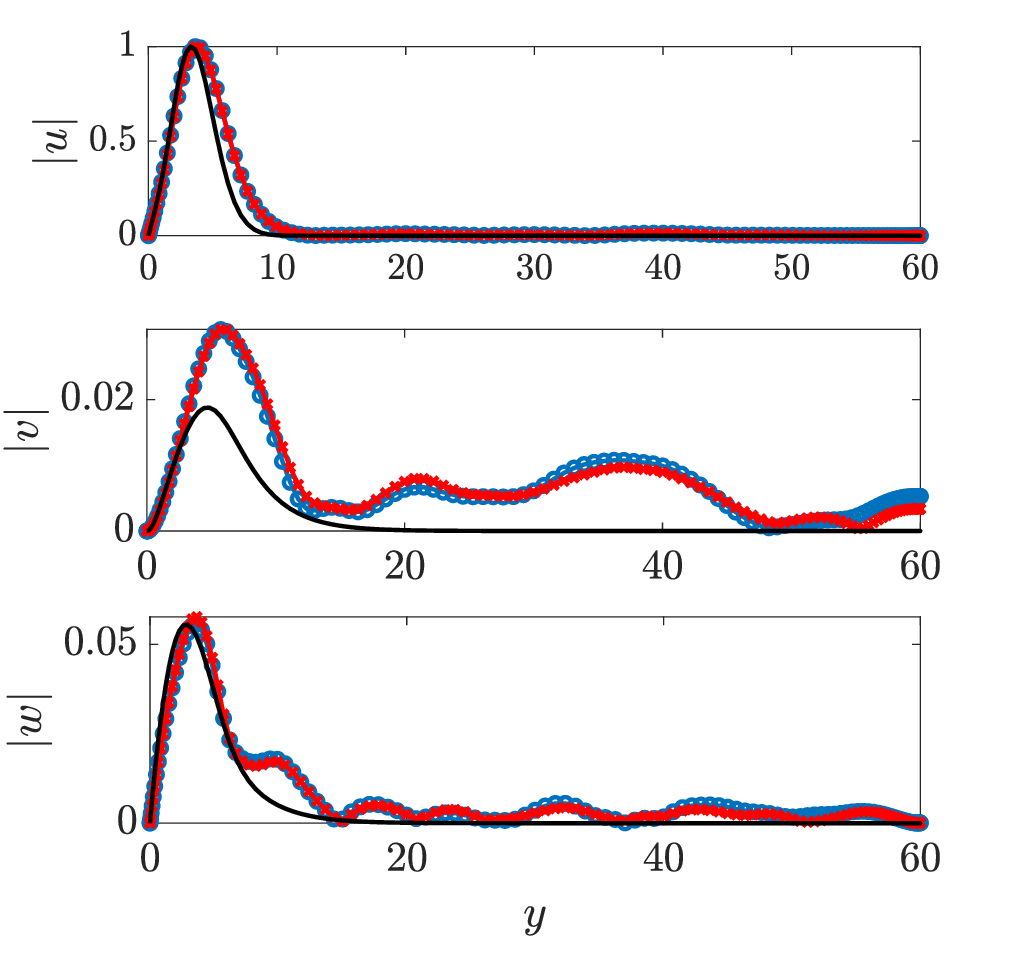}\\
		(a)
	\end{minipage}\hfill
	\begin{minipage}{0.5\textwidth}
		\centering
		\includegraphics[width=\linewidth]{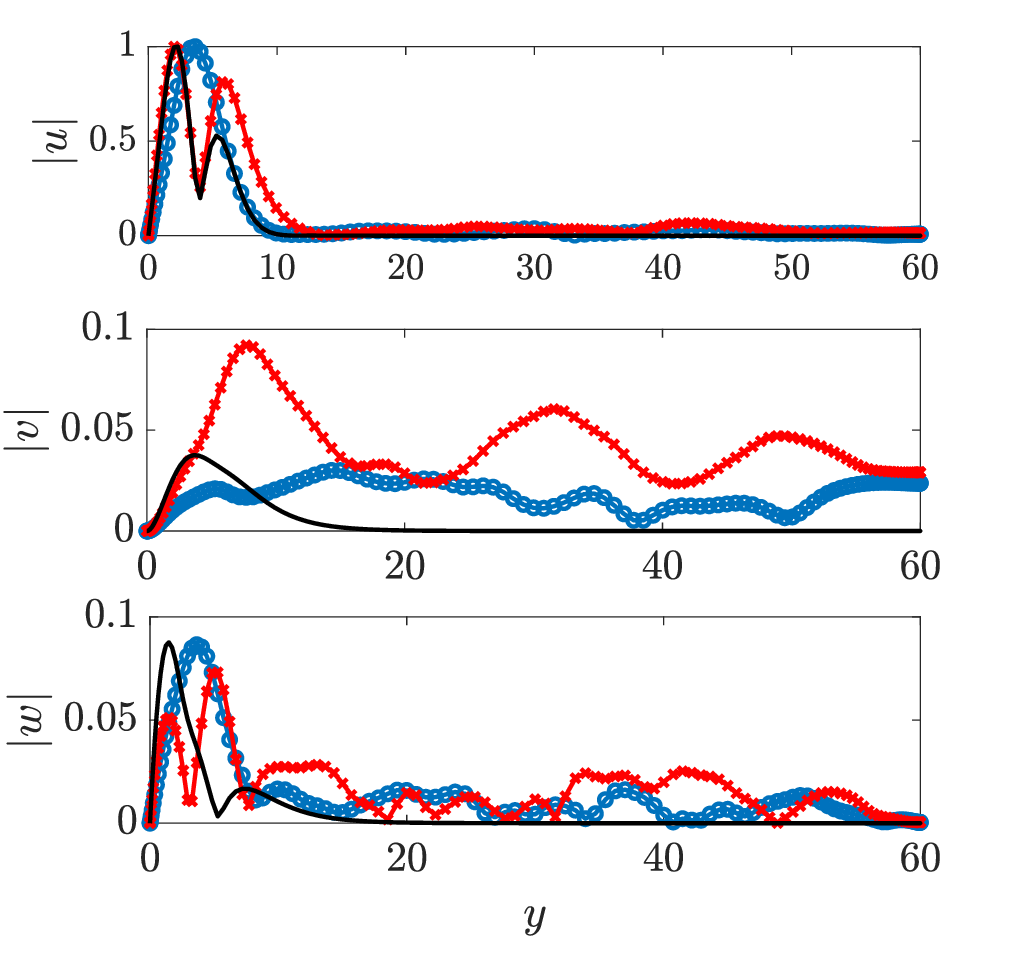}\\
		(b)
	\end{minipage}\hfill
	\caption{Normalised fluctuations at position $x=700$ for case 2. (a) Mode 1. (b) Mode 2. Legend: ($-$) Optimal linearised response (Resolvent) ($\color{blue}\circ$) Standard SPOD; ($\color{red}\times$) Shifted SPOD.}
	\label{fig:envelope_omega0}
\end{figure}

In a further analysis, focusing in case 2 ($N_{FFT}=192$), the absolute value of each of the mode components are plotted for the position $x=700$. In figure \ref{fig:envelope_omega0}a, the free-stream induced perturbations captured by SPOD modes are successful in matching the stream-wise component of the optimal linearised response predicted by the resolvent operator and the overall form of the transverse components, of much smaller magnitude and thus more difficult to converge, is consistent. This fact corroborates the assumption concerning the similarity between SPOD and Resolvent modes for large gain/energy separation, as is the case. It is also noted that leading shifted and non-shifted SPOD modes are practically identical at this stream-wise position, a characteristic well depicted in figure \ref{fig:quiver_omega0.0131}. To evaluate the differences between them a look into the subsequent mode is necessary (figure \ref{fig:envelope_omega0}b). For the second mode, the agreement between SPOD and resolvent is not as close. Nevertheless, the shifted mode better matches the two-peak form and the maximum magnitude position of the linearised stream-wise component.

To determine the statistical convergence of analysed modes, we apply the test employed in \cite{abreu_cavalieri_schlatter_vinuesa_henningson_2020} and compute SPOD modes considering 2 equal parts of the original database, each comprising 50\% of the total number of snapshots. Modes from part 1 are projected into modes from part 2 and alignment coefficients, defined in eq. (\ref{eq:alignment}) are calculated. For a given mode, a correlation close to 1 implies that the correspondent mode computed from the complete database is statistically converged. Results of this analysis (table \ref{tab:mu}) show that all modes achieve alignment coefficients higher than 90\%. This, in turn, indicates that discrepancies observed between shifted and non-shifted SPOD are not due to poor statistical significance, but rather to a reduction of bias caused by the temporal shift operation.

\begin{table}
	\centering
	\caption{Alignment coefficients at $N_{FFT}=192$, $\omega=0.0131$ and $\beta = 0.377$.}
	\begin{tabular}{|c|c|c|}
        \hline
        \rowcolor[HTML]{9B9B9B} 
        \textbf{}                             & \textbf{Mode 1} & \textbf{Mode 2} \\ \hline
        \rowcolor[HTML]{FFFFFF} 
        \cellcolor[HTML]{9B9B9B}Standard SPOD & 0.9637          & 0.9390          \\ \hline
        \cellcolor[HTML]{9B9B9B}Shifted SPOD  & 0.9842          & 0.9039          \\ \hline
    \end{tabular}
	\label{tab:mu}
\end{table}

\section{Conclusions}

In the present study we explore the effect of estimation parameters (number of blocks and number of snapshots per block) on the convergence of Spectral Proper Orthogonal Decomposition (SPOD), associated with the Welch method for spectral estimation. A convergence analysis is performed for a model linear complex Ginzburg-Landau system, forced with spatially uncorrelated signals, by measuring the estimation errors between numerical SPOD modes and resolvent response modes. 

Estimation errors in SPOD are found to be related to the spectral window length, normalised by the domain size and characteristic velocity. As seen in figures \ref{fig:errors} and \ref{fig:bias_noise}, two main phenomena affect the convergence of the SPOD method. For shorter windows, with duration comparable or lower than the flow-through time in the domain of interest, we average over a larger number of Welch blocks that are not sufficiently long to capture the space-time correlation in the system, which follows an advection velocity. This yields smooth albeit biased SPOD modes, meaning results are precise but not accurate. Inversely, for larger windows, we have fewer averaging blocks. This penalises precision as SPOD modes are contaminated with noise, due to residual variance.

An extension to the SPOD algorithm by means of a temporal data shift is proposed. By aligning cross-correlations between points inside each block, the new approach in shown to minimise the bias of SPOD modes computed with short window sizes, as seen in figure \ref{fig:errors_shifted}. This allows for better overall convergence as the SPOD can be computed with more blocks for a given time series. This is an important feature when dealing with large simulations that often count with a limited number of snapshots. Furthermore, when comparing the standard and shifted SPODs in a fixed-block framework of constant variance (figure \ref{fig:mode1_fixedBlocks}), shifted modes always display smaller estimation errors, achieving a order of magnitude difference for smaller windows.

In order to demonstrate the pertinence of the proposed method in fluid dynamics problems, we apply the temporal shift to an LES of a boundary layer subject to bypass transition, and SPOD modes are compared with global resolvent modes. Even though equivalence is no longer guaranteed for non-linear systems (coloured forcing), SPOD and resolvent modes are still similar when gain separation is large, as in the boundary layer case. 

Results show that the shifted SPOD, for the same estimation parameters, captures more energy in the leading mode when compared to the standard method. In SPOD modes, a spatial growth of stream-wise vortices is observed, a feature also present in the resolvent analysis but absent from spatial transient growth analysis \cite{doi:10.1063/1.869908,luchini2000}. This implies that the free-stream turbulence continuously "forces" stream-wise vortices, leading to additional spatial growth of disturbances. When a single stream-wise section is compared, shifted and non-shifted SPOD leading modes are very close and match the $u$ component of the optimal linear response (leading resolvent mode). Discrepancies arise when the second mode is looked upon. The shifted mode is closer to the $u$ component in the linear response. A further statistical convergence test indicates these changes are not due to lack of statistical significance, but rather to a reduction of bias caused by the shifting operation.

\appendix
\section{Relationship between SPOD and resolvent analysis for an arbitrary inner product norm}\label{appA}
Considering the inner-products
	\begin{equation}
		\left\langle \mathbf{\hat{y}}_{1}, \mathbf{\hat{y}}_{2}\right\rangle_{\mathbf{W_y}}=\mathbf{\hat{y}}_{1}^{H} \mathbf{W_y} \mathbf{\hat{y}}_{2}
	\end{equation}
	\begin{equation}
		\left\langle \mathbf{\hat{f}}_{1}, \mathbf{\hat{f}}_{2}\right\rangle_{\mathbf{W_f}}=\mathbf{\hat{f}}_{1}^{H} \mathbf{W_f} \mathbf{\hat{f}}_{2}
	\end{equation}
	and the resolvent operator $\mathbf{R}$, we compute response modes $\mathbf{U}$, forcing modes $\mathbf{V}$ and gains $\mathbf{\Sigma}$ via the SVD.
	\begin{equation}
		\mathbf{\tilde{y}} = \mathbf{W_y}^{1/2} \mathbf{\hat{y}}, \quad 
		\mathbf{\tilde{f}} = \mathbf{W_f}^{1/2} \mathbf{\hat{f}}
	\end{equation}
	\begin{equation}
		\mathbf{\tilde{y}} = \mathbf{\tilde{R}} \mathbf{\tilde{f}} \implies 
		\mathbf{\tilde{R}} = \mathbf{W_y}^{1/2} \mathbf{R} \mathbf{W_f}^{-1/2}
	\end{equation}
	\begin{equation}
		\mathbf{\tilde{R}} \mathbf{\tilde{V}} = \mathbf{\tilde{U}} \Sigma \implies \mathbf{R} \mathbf{W_f}^{-1/2} \mathbf{\tilde{V}} = \mathbf{W_y}^{-1/2} \mathbf{\tilde{U}} \Sigma
	\end{equation}
	\begin{equation} \label{eq:u_v}
		\mathbf{U} = \mathbf{W_y}^{-1/2} \mathbf{\tilde{U}}, \quad 
		\mathbf{V} = \mathbf{W_f}^{-1/2} \mathbf{\tilde{V}}
	\end{equation}

We assume
	\begin{equation}
		\mathbf{C}_\omega = \mathcal{E}\left(\mathbf{\hat{y}} \mathbf{\hat{y}}^H \mathbf{W_y}\right)
	\end{equation}
	\begin{equation}
		\mathbf{F}_\omega  = \mathcal{E}\left(\mathbf{\hat{f}} \mathbf{\hat{f}}^H \mathbf{W_f}\right) 
	\end{equation}
	which imply from the resolvent framework (eq. \ref{eq: resolvent_op}), 
	\begin{equation}
		\mathbf{C}_\omega = \mathbf{R} \mathbf{F}_\omega \mathbf{W_f}^{-1} \mathbf{R}^H \mathbf{W_y}
	\end{equation}

If forcing terms are perfectly uncorrelated, $\mathbf{F}_\omega=\mathbf{I}$ and the expression can be reduced to 
	\begin{equation}
		\begin{aligned}
			\mathbf{C}_\omega &= \mathbf{R} \mathbf{W_f}^{-1} \mathbf{R}^H \mathbf{W_y} = \mathbf{W_y}^{-1/2} \mathbf{\tilde{R}} \mathbf{\tilde{R}}^H \mathbf{W_y}^{1/2} \\
			&= \mathbf{W_y}^{-1/2} \mathbf{\tilde{U}} \Sigma^2 \mathbf{\tilde{U}}^{-1} \mathbf{W_y}^{1/2}
		\end{aligned}
	\end{equation}
	and thus, substituting eq. (\ref{eq:u_v}), we have
	\begin{equation}
		\mathbf{C}_\omega = \mathbf{U} \Sigma^2 \mathbf{U}^{-1}
	\end{equation}
	as in eq. (\ref{eq:res_spod}).

\section{On the convergence of the shifted spectral estimation}\label{appB}

Given two bounded, stationary signals $a(t)$ and $b(t)$, the Fourier transforms of the windowed signals within each Welch block are given by
	\begin{equation}
		\hat{a}_h(\omega) = \mathcal{F} \left\{h(t)a(t)\right\} = \frac{1}{T} \int_{t_0}^{t_0+T} h(t) a(t) e^{i \omega t} dt
	\end{equation}
	\begin{equation}
		\hat{b}_h(\omega,\tau) = \mathcal{F} \left\{h(t)b(t+\tau)\right\} =  \frac{1}{T} \int_{t_0}^{t_0+T} h(t) b(t+\tau) e^{i \omega t} dt
	\end{equation}
	where $t_0$ denotes the initial time for each block, $T$ the block duration, $h(t)$ the windowing function and $\tau$ the time shift.

Following the properties of the Fourier transform, a time shift $\tau$ generates a phase lag of $\exp \left( i \omega \tau\right)$. Thus, we write the corrected cross-periodogram as
	\begin{equation}
		S(\omega,\tau) = \hat{a}_h(\omega) \hat{b}_h(\omega,\tau) e^{-i \omega \tau}
	\end{equation}

In the limit where $T \to \infty$, we have
	\begin{equation}
		\lim_{T \to \infty} h(t) = 1
	\end{equation}
	\begin{equation}
		\lim_{T \to \infty} \hat{a}_h(\omega) = \mathcal{F} \left\{a(t)\right\}  = \hat{a}(\omega)
	\end{equation}
	\begin{equation}
		\lim_{T \to \infty} \hat{b}_h(\omega,\tau) = \mathcal{F} \left\{b(t+\tau)\right\}  = \hat{b}(\omega) e^{i \omega \tau}
	\end{equation}
	which leads to the true cross-periodogram
	\begin{equation}
		\lim_{T \to \infty} S(\omega,\tau) = \hat{a}(\omega) \hat{b}(\omega)
	\end{equation}
	and demonstrates that the procedure converges asymptotically for an arbitrary time shift and frequency. In that sense, shifted and non-shifted spectral estimations must return the same results for a large enough window size.  

\section*{Acknowledgments}
This work is financially supported by Fundação de Amparo à Pesquisa do Estado de São Paulo, FAPESP, under grants nos. 2019/27655-3 and 2020/14200-5 for Diego C. P. Blanco.

The scripts necessary to run the Ginzburg-Landau cases and SPOD computations are available in a public GIT repository at \url{https://github.com/DiegoCPB/Shifted-SPOD}.

\bibliographystyle{abbrv}
\bibliography{refs.bib}

\end{document}